\documentclass{aa}

\usepackage[tight]{subfigure}
\usepackage{graphicx}
\usepackage{txfonts}
\usepackage{color}

\usepackage{chemarrow}

\usepackage{natbib}
\bibpunct{(}{)}{;}{a}{}{,}

\begin{document}
  \title{The interaction of core-collapse supernova ejecta with a companion star}


   \author{Zheng-Wei Liu
          \inst{1},
         T. M. Tauris 
          \inst{1,2},
          F. K. R\"opke
          \inst{3,4},
         T. J. Moriya
          \inst{1},
          M. Kruckow
          \inst{1},
          R. J. Stancliffe
          \inst{1},
          \and
          R. G. Izzard
          \inst{1,5}
                    }

   \institute{Argelander-Institut f\"ur Astronomie, Universit\"at Bonn, Auf dem H\"ugel 71, 53121 Bonn, Germany\\
            \email{zwliu@ynao.ac.cn}
         \and
              Max-Planck-Institut f\"ur Radioastronomie, Auf dem H\"ugel 69, D-53121 Bonn, Germany\
         \and
             Heidelberger Institut f\"ur Theoretische Studien, Schloss-Wolfsbrunnenweg 35, D-69118 Heidelberg, Germany\
         \and
             Zentrum f\"ur Astronomie der Universit\"at Heidelberg, Institut f\"ur Theoretische Astrophysik, Philosophenweg 12, D-69120 Heidelberg, Germany\
         \and
            Institute of Astronomy, University of Cambridge, Madingley Road, Cambridge CB3 0HA, United Kingdom
           }

   \date{Received June 16, 2015; accepted September 10, 2015}

 \abstract
   { The progenitors of many core-collapse supernovae (CCSNe) are expected to be in binary systems. After the SN explosion 
     in a binary, the companion star may suffer from mass stripping and be shock heated as a result of the impact of the
     SN ejecta. If the binary system is disrupted by the SN explosion, the companion star is ejected as a runaway 
     star, and in some cases as a hypervelocity star.}
   {By performing a series of three-dimensional (3D) hydrodynamical simulations of the collision of SN ejecta with 
    the companion star, we investigate how CCSN explosions affect their binary companion.}
   {We use the {\sc BEC} stellar evolution code to construct the detailed companion structure at the moment of SN explosion. 
    The impact of the SN blast wave on the companion star is followed by means of
    3D smoothed particle hydrodynamics (SPH) simulations using the {\sc Stellar GADGET} code.
}
   { For main-sequence (MS) companion stars, we find that the amount of removed stellar mass, the resulting impact velocity, and
     the chemical contamination of the companion that results from the impact of the SN ejecta strongly increases with decreasing 
     binary separation and increasing explosion energy. Their relationship can be approximately fitted by power laws, 
     which is consistent with the results obtained from impact simulations of Type~Ia~SNe.  
     However, we find that the impact velocity is sensitive to the momentum profile of the outer SN ejecta and, in fact,
     may decrease with increasing ejecta mass, depending on the modeling of the ejecta.
     Because most companion stars to Type~Ib/c CCSNe are in their MS phase at the moment of the explosion, combined with the
     strongly decaying impact effects with increasing binary separation, we argue that the majority of these SNe lead to inefficient mass stripping 
     and shock heating of the companion star following the impact of the ejecta. }
   {Our simulations show that the impact effects of Type Ib/c SN ejecta on the structure of MS companion stars, 
    and thus their long-term post-explosion evolution, is in general not dramatic. We find that at most
    10\% of their mass is lost and their resulting impact velocities are less than 100~${\rm km\,s}^{-1}$. 
 }

                
   \authorrunning{Z. W. Liu et al.}
   
   \titlerunning{The interaction of CCSN ejecta with a companion star}

   \maketitle
%

\section{Introduction}\label{sec:introduction}
The discovery of many low-mass X-ray binaries and millisecond pulsars in tight orbits, i.e., binary 
neutron stars with orbital periods of less than a few hours, provides evidence 
for supernova (SN) explosions in close binaries with low-mass companions.
The nature of the SN explosion determines whether any given binary system remains bound or is disrupted \citep{Hills83}. 
An additional consequence of the SN explosion is that the companion star is affected by the impact of the 
shell debris ejected from the exploding star \citep{Whee75}. 
Besides chemical enrichment, this kind of an impact has kinematic effects and may induce 
significant mass loss and heating of the companion star.

Core-collapse supernovae (CCSNe) arise from massive stars. Determining the relationship between SNe 
and their progenitor stars remains an important problem (see \citealt{Lang12}, for a recent review). 
Despite the fact that the amount of SN ejecta is decisive for the aftermath of a SN in a binary system, 
the mechanism of significant, or even complete, removal of the hydrogen~(H) envelope in H-deficient CCSNe 
(Type~Ib/c, IIb, and IIL) is not well understood. 

Although it was originally proposed that H-deficient SNe might arise from massive, metal-rich 
single stars through stellar wind mass loss \citep{Woos93,Maed03,Eldr04}, it has long been 
considered that mass transfer via Roche-lobe overflow (RLO) in binary systems is a likely mechanism to produce the 
progenitors of H-deficient CCSNe (\citealt{Pacz67,Pods92,Pods93,Maun04,Eldr08, Stan09, Yoon10,Dess11}).
There is growing observational evidence that the fraction of massive stars 
in close binary systems is large \citep{chn+12,Duch13}. 
\citet{Sana12} found that more than $70\%$ of massive stars are in close binary systems, which supports the idea that 
binary progenitors contribute significantly to the observed H-deficient SNe. 

Evidence from statistical samples of Type~Ib/c~SNe suggests that stripped, relatively low-mass, 
exploding stars in binaries are likely to be progenitors of most Type~Ib/c~SNe \citep{Drou11,Bian14,Lyma14,Tadd15}.
In some cases, like SN 2005ek, there is even evidence of extreme stripping of the SN progenitor star, with only 
$\sim\!0.1\,M_{\odot}$ of material being ejected \citep{dsm+13,tlm+13,tlp15}.  
With $\textit{HST}$ analysis of pre-SN images, \citet{Cao13} report the detection of a possible progenitor candidate 
at the location of a young, H-deficient Type~Ib~SN, iPTF13bvn\footnote{This may be the first direct detection of a progenitor of a 
Type~Ib~SN.}. Further light-curve modeling, and analysis of pre-SN and late-time photometry, of iPTF13bvn suggests 
that an interacting binary system is likely to be the progenitor of iPTF13bvn rather than a 
standard single Wolf-Rayet star (\citealt{Bers14,Frem14,Eldr15}, but see \citealt{Groh13}). All these findings indicate that H-deficient CCSNe 
and in particular Type~Ib/c~SNe, can naturally be produced in binary progenitor systems. 

In the binary progenitor scenario of Type~Ib/c~CCSNe, the metal core of a massive star (originally $8$--$30\,M_{\odot}$) 
undergoes gravitational collapse to form a neutron star (NS), while the outer parts of its envelope are expelled 
at a high velocity and impact on the binary companion star. Depending on the amount of sudden mass loss because of 
the SN explosion and the momentum kick imparted on the newborn NS, the outcome of the SN explosion is 
either a bound system \citep{Hills83}, which may later evolve into an X-ray binary system \citep{TvdH06}, or two single 
runaway stellar components, which are left behind if the binary is unbound \citep{Taur98}. 
This kind of a disruption of close binaries by a SN explosion has been proposed as a possible mechanism for producing
hypervelocity stars (HVSs) that will leave our Galaxy because both ejected stars can receive high runaway velocities 
when asymmetric SNe are considered \citep{Taur98,Taur15}.  

After a SN explosion occurs in a binary system, the ejected debris is expected to expand freely for a 
few minutes to hours and eventually impact the companion star. As a result, the early supernova light curve can be brightened by
the collision of the supernova ejecta with the companion star \citep{Kase10, Mori15, Kuts15, Liu15}. Also, the companion star may be 
significantly heated and shocked by the SN impact, causing the envelope of the companion star to be  
partially removed due to the stripping and ablation mechanism \citep{Whee75, fa81, tf84}. Many numerical simulations 
of collisions of SN ejecta with the companion star in single-degenerate Type~Ia~SNe have 
been carried out by different groups \citep{Whee75,tf84,Mari00,Pakm08,Liu12,Liu13b,Pan10,Pan12}. All 
these studies suggest that about $2$--$30\%$ of the companion mass, or almost the whole envelope of the companion 
star (depending on its structure), is removed by the SN impact. 
Nondegenerate companion stars are expected to survive the impact of the SN explosion and show some peculiar properties,
such as high runaway velocities, overluminosity, and probably enrichment in heavy elements, which may be identified by 
observations. 

Whereas simulations of Type~Ia~SNe are generally performed based on a fixed (or small range of) explosion energy, $E_{\rm ej}$,
and a fixed ejecta mass, $M_{\rm ej}$, CCSNe can have a large spread in $E_{\rm ej}$ and $M_{\rm ej}$.  
\citet{Hira14} carried out two-dimensional impact simulations to investigate  
the collision of CCSN ejecta with a $10\,M_{\odot}$ giant star companion. However, most companion stars to CCSNe 
are on the main sequence (MS) at the moment of first SN explosion in a binary system (see \citealt{Koch09}, and also our 
population synthesis calculations in Section~\ref{sec:bps}).

In this work, we perform impact simulations using a three-dimensional (3D) smoothed particle hydrodynamics (SPH) 
method to systematically study, for the first time, the impact of CCSN ejecta on MS companion stars.
In Section~\ref{sec:code}, we describe the methods and initial models used in this work. Our 
impact simulations are presented in Section~\ref{sec:results}, and a parameter survey and a detailed discussion 
follow in Section~\ref{sec:discussion}. Our conclusions are summarized in Section~\ref{sec:summary}.

\section{Numerical methods and models}\label{sec:code}
In this section, we introduce the codes and methods used in this work, and we construct the 
initial companion star models and SN ejecta models. Also, initial
conditions and setup of our hydrodynamical impact simulations are described in detail.

\subsection{Numerical codes}
To construct the detailed structure of the companion star at the 
moment of the CCSN explosion, we use the one-dimensional (1D) implicit 
Lagrangian stellar evolution code {\sc BEC} \citep{Lang91,Lang98, Yoon06, ydl12}, which solves 
the hydrodynamic stellar structure and evolution equations \citep{Kipp90}. 
A recent description of the updated code is found in \citet{Yoon06, Yoon10}. 
In our default models we assume a mixing-length parameter 
$\alpha=l/H_{\rm{P}}=1.5$ \citep{Lang91} and a core convective overshooting parameter 
of $\delta_{\rm{ov}}=0.10$. We evolve stars with a given mass and solar metallicity ($Z=0.02$)  
to construct 1D stellar models for our subsequent impact simulations.

In our hydrodynamical simulation of the collision of the SN ejecta on the binary companion 
star, we use the 3D smoothed particles hydrodynamics (SPH) code {\sc Stellar GADGET} 
\citep{Pakm12}. An advantage of applying SPH for the problem under investigation
is that no material leaves the computational domain, and also that
momentum, energy, and angular momentum are strictly conserved.
The {\sc GADGET} code \citep{Spri05} was originally intended for cosmological simulations, 
but it has been modified by \citet{Pakm08,Pakm12} for stellar astrophysics problems \citep{Pakm08,Liu12}. 
As mentioned above, 
several studies have successfully used the {\sc Stellar GADGET} code to investigate the 
SN impact on the companion star in single-degenerate SNe~Ia \citep{Pakm08,Liu12,Liu13a,Liu13b,Liu13c}. 
In particular, \citet{Pakm08} have shown that the SPH-based approach reproduces previous 
results obtained with a grid-based 2D scheme by \citet{Mari00}. This shows that the
SPH approach with the {\sc Stellar GADGET} code captures the main dynamical effects of the collision 
between SN ejecta and its companion star.

\subsection{Binary companion star models}
\begin{figure}
  \begin{center}
    {\includegraphics[width=0.95\columnwidth, angle=360]{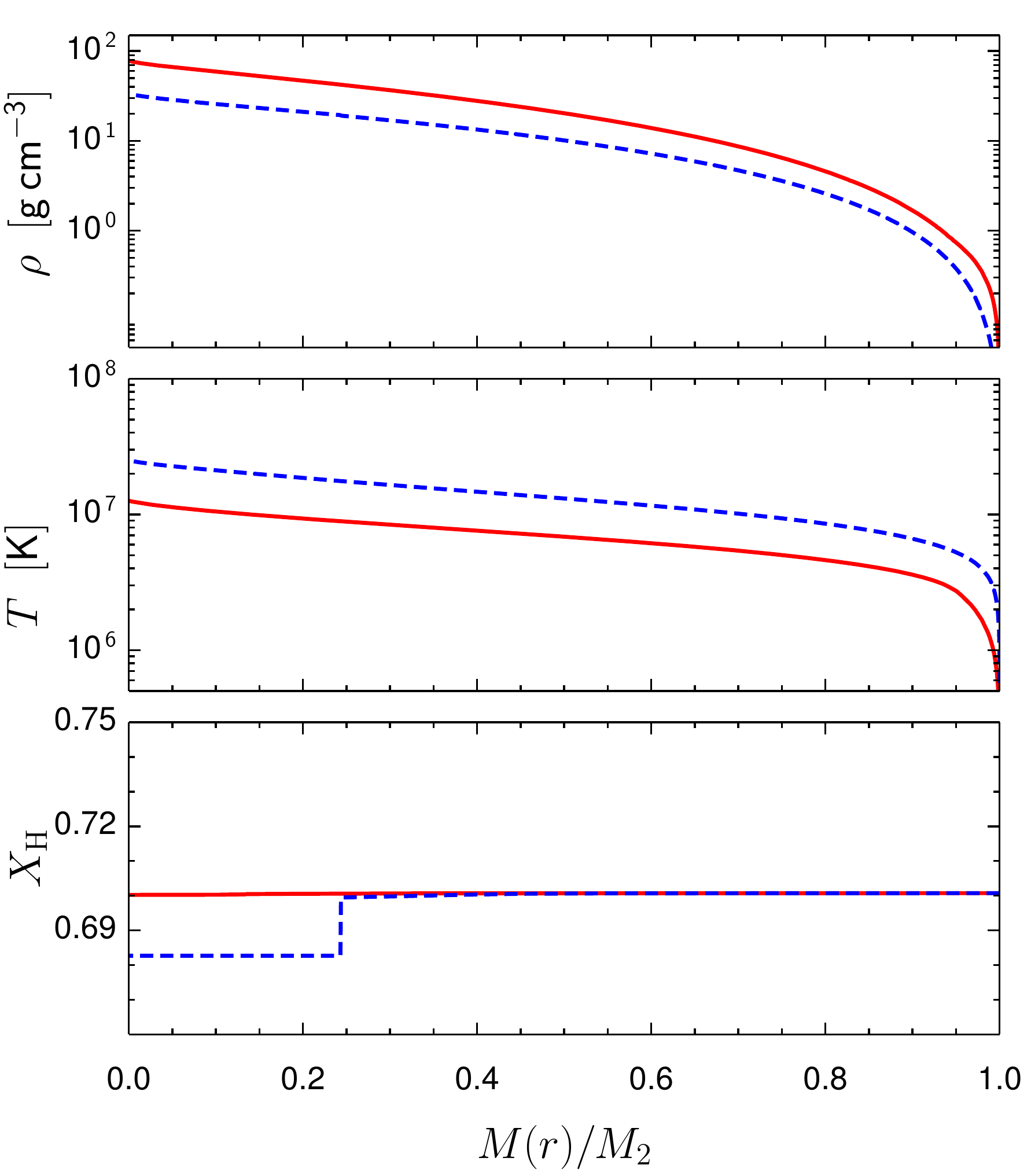}}
  \caption{ Profiles of mass density $\rho$, temperature $T$, and hydrogen abundance $X_{\rm{H}}$ as a
            function of $M(r)/M_{2}$ for a $0.9\,M_{\odot}$ G/K-dwarf (red solid lines) 
            and a $3.5\,M_{\odot}$ late-type B-star (blue dashed lines) companion star model 
            at the moment of the CCSN explosion. Here, $M(r)$ is the mass enclosed within radius $r$, $M_{2}$
            is the total mass. 
             }
\label{Fig:1}
  \end{center}
\end{figure}

The maximum possible runaway velocities for MS HVSs ejected from disrupted binaries in
asymmetric SN explosions have been investigated by \citet{Taur15}. 
In their systematic Monte Carlo simulations of SN-induced HVSs, they chose to focus on MS companion stars with masses 
$M_{2}=0.9\,M_{\odot}$ and $3.5\,M_{\odot}$. 
This selection is based on observations of the main class of HVSs \citep[G/K-dwarf,][]{bgkk05,bgk09,bgk12,bgk14}
and the recently discovered late-type B-star HVS candidates \citep{psh+14}.   
In this work, the same two companion stars used in \citet{Taur15} are used in our impact hydrodynamical simulations. 
Using BEC, we evolved zero-age main-sequence (ZAMS) stars of these two masses to an age of about 6~Myr, roughly the 
lower limit of their expected age when their massive companion star explodes.
The structure profiles of these two companion stars at the moment of the SN explosion are shown in Fig.~\ref{Fig:1}.

\subsection{The supernova model}
To introduce a SN explosion, we adopt a simple analytical explosion ejecta model which is constructed based on
numerical simulations of SN explosions by \citet{Matz99}. We assume that the SN ejecta is already in homologous 
expansion, whereby the radius of a fluid element is given by $r=v_{\rm{ej}}\,t$, where $v_{\rm{ej}}$ is the expansion velocity, and $t$ is the time 
since the explosion. The density profile of the expanding SN ejecta, $\rho_{\rm{ej}}(v_{\rm{ej}},t)$, is 
described by a broken power law, $\rho_{\rm{ej}}\propto r^{-n}$ for the outer part of the SN ejecta and $\rho_{\rm{ej}}\propto r^{-\delta}$ 
for the inner part of the SN ejecta, as follows (\citealt{Kase10, Mori13}):
 \begin{equation}
    \label{eq:1}
\small
\rho_{\rm{ej}}(v_{\rm{ej}},t) = \left\{ \begin{array}{lll}
\displaystyle \frac{1}{4\pi(n-\delta)} \frac{[2(5-\delta)(n-5)\,E_{\rm{ej}}]^{(n-3)/2}}{[(3-\delta)(n-3)\,M_{\rm{ej}}]^{(n-5)/2}}\,t^{-3}\,v_{\rm{ej}}^{-n}   & (v_{\rm{ej}}>v_{\rm{t}}),  \\
 & & \\
\displaystyle \frac{1}{4\pi(n-\delta)} \frac{[2(5-\delta)(n-5)\,E_{\rm{ej}}]^{(\delta-3)/2}}{[(3-\delta)(n-3)\,M_{\rm{ej}}]^{(\delta-5)/2}}\,t^{-3}\,v_{\rm{ej}}^{-\delta}   & (v_{\rm{ej}}<v_{\rm{t}}),
\end{array} \right.
  \end{equation}
where $E_{\rm{ej}}$ and $M_{\rm{ej}}$ are the SN explosion energy and SN ejecta mass, respectively, and 
 \begin{equation}
    \label{eq:2}
v_{\rm{t}} =\left [\frac{2(5-\delta)(n-5)\,E_{\rm{ej}}}{(3-\delta)(n-3)\,M_{\rm{ej}}}\right]^{1/2}. 
  \end{equation}

In SNe~Ib/c, typically $n\simeq 10$ and $\delta \simeq 1$ (\citealt{Matz99}), and we adopt these to 
construct a homologously expanding SN ejecta model based on the above Eqs.~(\ref{eq:1}) and~(\ref{eq:2}).  
Ideally one should start from a detailed simulation of the collapse of the exploding star, however, we limit ourselves 
to exploring the basic processes during the SN impact. A more realistic SN explosion model from hydrodynamical
simulations may slightly change our numerical results. However, detailed 
parameter surveys are addressed in Section~\ref{sec:discussion} to mimic different SN ejecta 
models, and, therefore, we do not expect a significant change in our main conclusions.

The CCSN explosion energy ($E_{\rm ej}=1.0-8.0\times 10^{51}\;{\rm erg}$), the amount of mass ejected ($M_{\rm ej}=0.7-5.0\;M_{\odot}$),
and the pre-SN orbital separation ($a=1.0-6.0\;a_{\rm min}$) are treated as free parameters ($a_{\rm min}$ is defined below).

\subsection{Initial setup}
Our basic setup and initial conditions for the impact simulations are similar to  
those of \citet{Liu12,Liu13a,Liu13c}. We use the healpix method described in \citet{Pakm12} to map the  
profiles of density and internal energy of a 1D companion star model to a particle distribution 
suitable for the SPH code (\citealt{Liu13c}). Before we start the actual impact simulations, 
the SPH model of each companion star is relaxed for ten dynamical timescales 
($t_{\rm{dyn}}=(G\rho)^{-1/2}/2$) to reduce numerical noise introduced by the mapping. If the relaxation 
succeeds, the velocities of the particles stay close to zero.
Otherwise, we reject the SPH model and redo the relaxation after adjusting the relaxation parameters \citep{Pakm12}.
We use $6\times10^6$ SPH particles to represent the B-type star or G/K-dwarf companion star.

The SN ejecta model described above is placed at a distance larger than $a_{\rm{min}}$ from the companion star in 
the x-y plane. A star in a binary cannot be closer to its companion than the Roche-lobe radius because mass 
transfer would result. We therefore use the Roche-lobe radius of MS companion star \citep{Eggl83} to set the minimum 
possible separation $a_{\rm{min}}$ in our systems (\citealt{Taur15}), 
\begin{equation}
    \label{eq:3}
 a_{\rm{min}}=\frac{0.6\,q^{2/3}+\rm{ln}(1+\textit{q}^{1/3})}{0.49\,q^{2/3}}\,R_{2}\  , 
  \end{equation}
where $q$ is the mass ratio between the companion star ($M_2$) and the exploding star ($M_{\rm He}$), i.e. 
$q=M_2/M_{\rm{He}}$, and $R_{2}$ is the radius of the companion star \citep{Eggl83}. 
As a consequence, the minimum orbital separations between the exploding star and our companion stars are 
$2.74\,R_{\rm{\odot}}$ (G/K-dwarf companion) and $5.56\,R_{\rm{\odot}}$ (late-type 
B-star companion), corresponding to $a_{\rm min}=3.55\;R_2$ and $a_{\rm min}=2.55\;R_2$, respectively. 
We assume that the exploding star is an evolved helium star of a mass $M_{\rm He}=3.0\;M_{\odot}$ which corresponds to a SN ejecta
mass of $\approx\! 1.4\;M_{\odot}$ with the formation of a NS with a gravitational mass of $\approx\! 1.4\;M_{\odot}$,
given that about $0.2\;M_{\odot}$ is released by neutrinos from lost gravitational binding energy during the explosion.   
The impact of the SN ejecta on the binary companion is simulated for more than $\rm{6000\,s}$, at 
which point the amount of mass removed from the companion envelope and its resulting impact 
velocity have converged (Fig.~\ref{Fig:2}).

\section{Hydrodynamical results}\label{sec:results}
In this section, our impact simulations of a G/K-dwarf companion star model ($M_{2}=0.9\,M_{\odot}$) with an 
initial binary separation of $a=5.48\,R_{\odot}$ ($\approx 7.1\;R_2\approx\! 2.0\ a_{\rm{min}}$) are chosen as an example. The detailed 
evolution of the companion star during the impact of the SN ejecta is qualitatively described. Under the assumption of 
homologous expansion, a typical SN ejecta model in our simulation is constructed according to 
Eqs.~(\ref{eq:1}) and (\ref{eq:2}), with $E_{\rm{ej}}=1.0\times10^{51}\,\rm{erg}$  and $M_{\rm{ej}}=1.4\,M_{\odot}$ . 
The maximum velocity at the ejecta front is set large enough ($\approx\! 30\,000\,\rm{km\,s^{-1}}$) that the 
assumed $E_{\rm{ej}}$ is actually contained in the ejecta.

\begin{table}
\caption{Resolution test for a G/K-dwarf companion model.}
\label{table:1}
\centering
\begin{tabular}{r r c c c }     
\hline\hline
          &  & $\Delta M_{\rm{2}}$ &$v_{\rm{im}}$ & $\Delta M_{\rm{acc}}$ \\
$N_{\rm{star}}$ &$N_{\rm{total}}$ &[$M_{\rm{\odot}}$]&[$\rm{km\ s^{-1}}$] &[$M_{\rm{\odot}}$]\\
\hline
 5\ 000       & 12\ 640      & 0.0327 & 60.4 & $3.7\times10^{-4}$\\
 10\ 000      & 25\ 427      & 0.0256 & 55.8 & $8.1\times10^{-4}$\\
 50\ 000      & 127\ 656     & 0.0123 & 37.8 & $8.8\times10^{-4}$\\
 100\ 000     & 255\ 513     & 0.0105 & 33.4 & $9.3\times10^{-4}$\\
 500\ 000     & 1\ 278\ 568  & 0.0066 & 27.4 & $9.5\times10^{-4}$\\
 1\ 000\ 000  & 2\ 556\ 994  & 0.0062 & 24.8 & $9.7\times10^{-4}$\\
 2\ 000\ 000  & 5\ 113\ 998  & 0.0057 & 22.6 & $9.6\times10^{-4}$\\
 4\ 000\ 000  & 10\ 227\ 949 & 0.0055 & 22.5 & $9.8\times10^{-4}$\\
 6\ 000\ 000  & 15\ 341\ 986 & 0.0055 & 22.6 & $9.7\times10^{-4}$\\

\hline
\end{tabular}

\tablefoot{ $N_{\rm{star}}$ and $N_{\rm{total}}$ are the number of particles 
  used to represent the companion star and the binary system, respectively.
 All particles have
 the same mass $M_{\rm{SPH}}$. 
}
         
\end{table}

\begin{figure}
  \begin{center}
    {\includegraphics[width=0.95\columnwidth, angle=360]{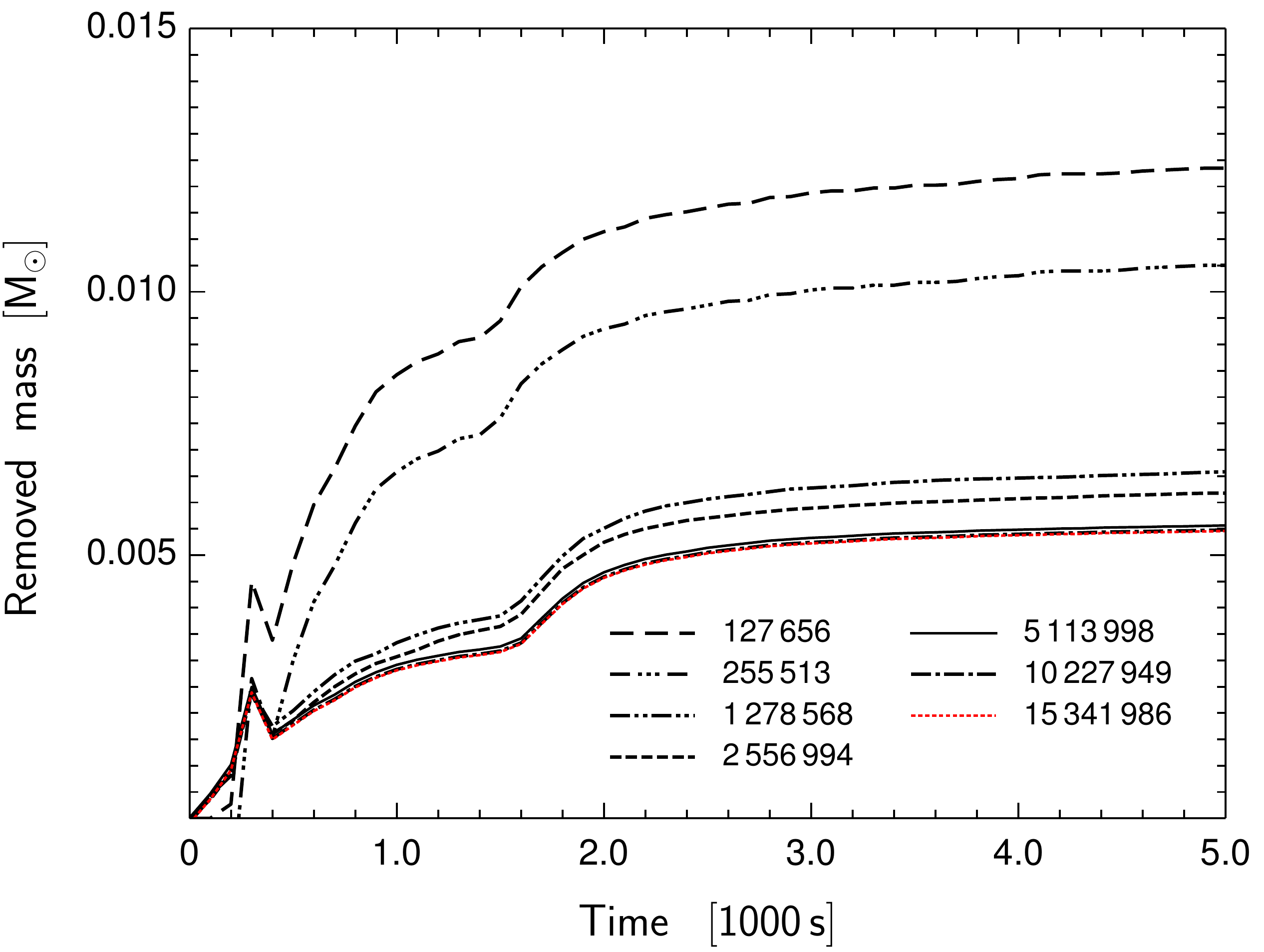}}
  \caption{ Amount of mass removed from the companion star as a function of time in the
            impact simulation of a G/K-dwarf companion ($M_{2}=0.9\,M_{\odot}$, $a=5.48\,R_{\odot}\approx 7.1\;R_2$) 
            with different resolutions, i.e., different total numbers of SPH particles in the simulation (Table~\ref{table:1}).
            The standard resolution used in our simulation is highlighted in red.                
             }
\label{Fig:resolution}
  \end{center}
\end{figure}

\subsection{Resolution test}

To assess the reliability of our numerical results, we have performed a convergence test to check 
the sensitivity of our results to different resolutions by varying the particle numbers
from $5\,000$ to $6\times10^{6}$ in the companion star (Table~\ref{table:1}). Because all our SPH particles have 
the same mass, the number of SPH particles representing the SN ejecta is fixed 
once the SPH companion model is created.

\begin{figure}
  \begin{center}
    {\includegraphics[width=0.93\columnwidth, angle=360]{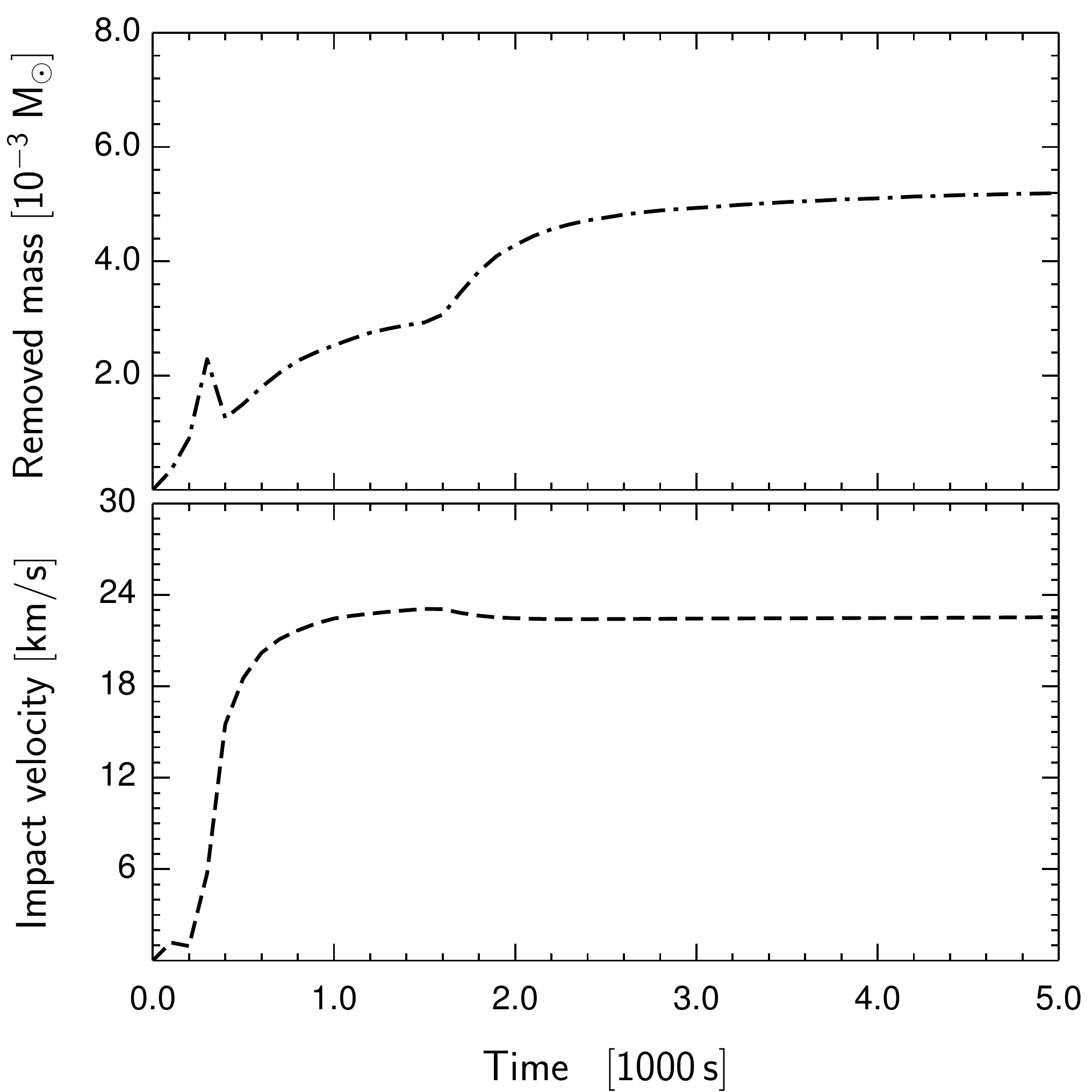}}
  \caption{ Amount of mass removed from the companion star (top panel) and the impact velocity caused by the SN 
            ejecta (bottom panel) as a function of time after the
            SN explosion in our impact simulations of a G/K-dwarf 
            companion ($M_{2}=0.9\,M_{\odot}$) with an initial binary 
            separation of $a=5.48\,R_{\odot}\approx 7.1\;R_2\approx 2.0\;a_{\rm min}$. Here, $a_{\rm min}$
            is the minimum possible separation of the binary system.}
\label{Fig:2}
  \end{center}
\end{figure}

\begin{figure*}
  \begin{center}
    {\includegraphics[width=2.0\columnwidth, angle=360]{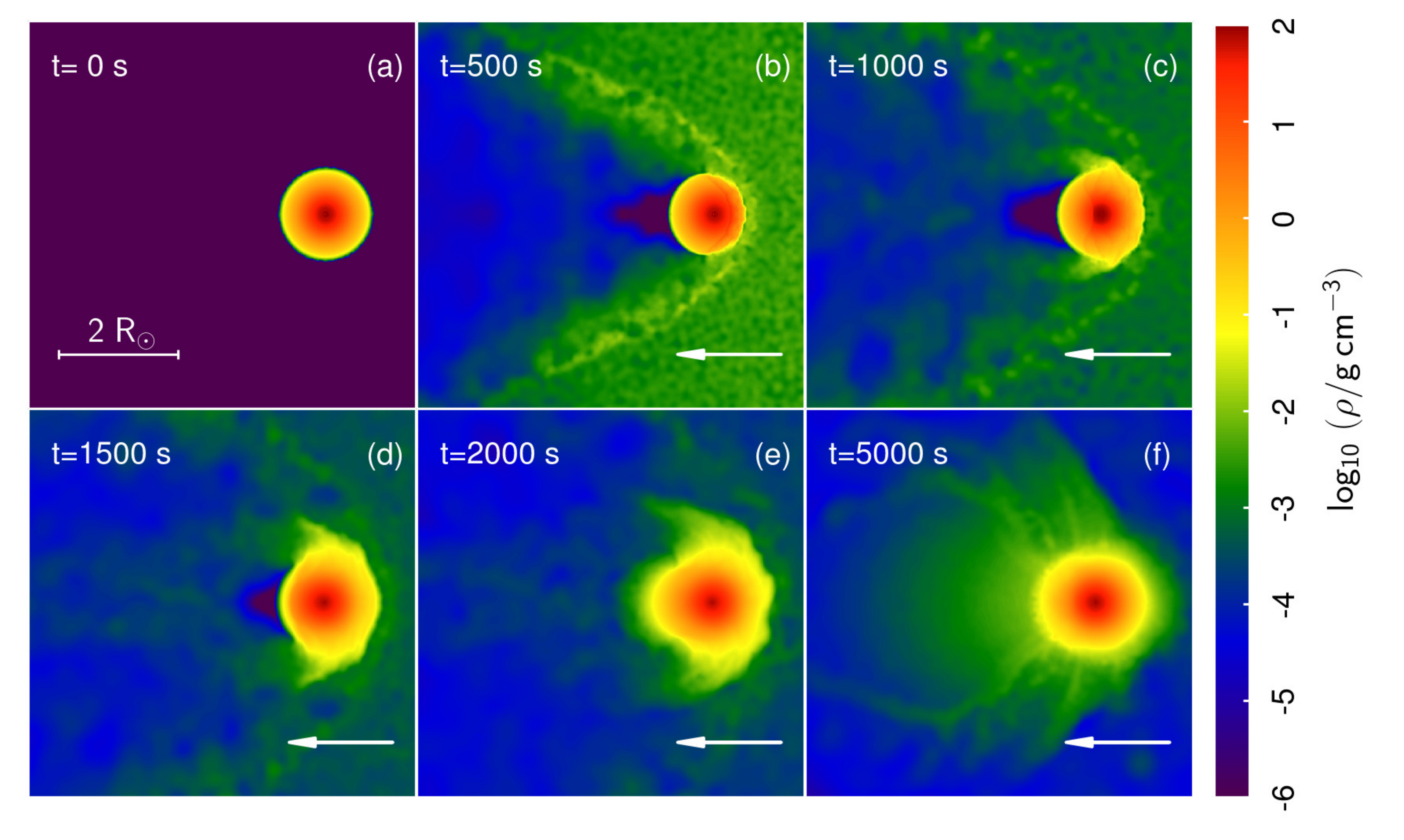}}
  \caption{Density distributions of all gas material as a function of the explosion time in the impact simulations for a G/K-dwarf 
          companion model ($M_{2}=0.9\,M_{\odot}$) with a binary separation of $5.48\,R_{\odot}$ ($=7.1\;R_2$). The direction of motion of the incoming SN shell front is 
          from right to left (see arrow symbols). The impact is initiated at $t\approx 100\;{\rm s}$ after the explosion.
          The color scale shows the logarithm of the mass density in $\rm{g\,cm^{-3}}$ . }
\label{Fig:3}
  \end{center}
\end{figure*}

The amount of removed companion mass as a function of time in our impact simulations on a 
G/K-dwarf companion star model ($a=5.48\,R_{\odot}$) with different resolutions is shown in Fig.~\ref{Fig:resolution}.
To calculate the amount of mass removed (evaporated) from the companion star by the SN impact, we 
sum over the masses of all unbound SPH particles (stripped+ablated particles) that originally belonged to it. 
Whether or not a particle is bound to the star is determined by calculating its total energy, 
$E_{\mathrm{tot}}= E_{\mathrm{kin}}+ E_{\mathrm{pot}} + E_{\mathrm{in}}$, where $E_{\mathrm{kin}}$, $E_{\mathrm{pot}}$ 
and $E_{\mathrm{in}}$ are the kinetic energy (positive), potential energy (negative) and 
internal energy (positive), respectively. If $E_{\mathrm{tot}} > 0$, the particle is unbound. The center-of-mass motion of the star is subtracted when 
calculating the kinetic energy of each particle. 
To make the difference between the results obtained with particle numbers above 2 million better visible, 
only the results
with a resolution ranging from about $1.3\times10^{5}$ to $1.5\times10^{7}$ SPH particles are plotted in Fig.~\ref{Fig:resolution}.
The difference in removed companion mass between simulations with $5.1\times10^{6}$ (the solid line) and $1.0\times10^{7}$ (the dash-dotted line) 
SPH particles is smaller than $4\%$, the difference between $1.0\times10^{7}$ and $1.5\times10^{7}$ (the dotted line) SPH 
particles is less than $2\%$ (Table~\ref{table:1}). We find that the amount of removed companion mass ($\Delta M_{2}$), 
the impact velocity ($v_{\rm{im}}$) and the amount of accreted ejecta mass ($\Delta M_{\rm{acc}}$) 
are sufficiently converged (Table~\ref{table:1}) when using more than $4\times10^6$  SPH particles 
to represent the companion star and we thus settle on the number $6\times10^6$  (which corresponds to a 
total of about $10^7$  particles in the simulation). The mass of a single particle, $M_{\rm{SPH}}$, is then $\approx 10^{-6}-10^{-7}\,M_{\odot}$.

\subsection{Description of evolution during/after SN ejecta impact}\label{sec:density}
Figure~\ref{Fig:3} demonstrates the typical evolution of the gas density in the orbital plane 
in our impact simulation of a $0.9\ M_{\odot}$ G/K-dwarf companion model with an initial 
binary separation of $5.48\ R_{\odot}\approx 7.1\;R_2$.  The basic impact processes are quite similar 
to previous hydrodynamical simulations with a MS donor model in normal SNe Ia \citep{Liu12,Pan12}.

At the beginning of the simulation, the MS companion star is in equilibrium (Fig.~\ref{Fig:3}a). The SN explodes on the right side of the companion 
star, the SN ejecta expands freely for a while and hits the MS companion star, 
removing H-rich material from its surface through stripping (momentum transfer) and
ablation (SN shock heating). The latter is caused by a bow shock
propagating through the star, leading to internal heating and additional loss of mass from the far side of the companion star (Fig.~\ref{Fig:3}e). 
In the final snapshot ($t=5000\;{\rm s}$), all removed companion material caused by the SN impact is 
largely embedded in a low-velocity SN debris behind the companion star. At the end of the simulation,  
we calculate the amount of removed companion mass, $\Delta M_2\approx\! 0.0055\,M_{\odot}$. Mass loss caused 
by the SN impact reaches a steady state by the end of the simulation, we therefore do not expect 
any further loss (Fig.~\ref{Fig:2}). In Fig.~\ref{Fig:2}, a small peak of the removed mass is seen 
in the early phase after the SN ejecta hits the companion star. This is because a shocked particle in the center
of the star can have a larger kinetic than potential energy when the shock wave passes through
the star. However, it is not able to leave the star and soon loses its
kinetic energy to some neighbor particles and therefore becomes bound again.

\begin{figure}
  \begin{center}
    {\includegraphics[width=1.0\columnwidth, angle=360]{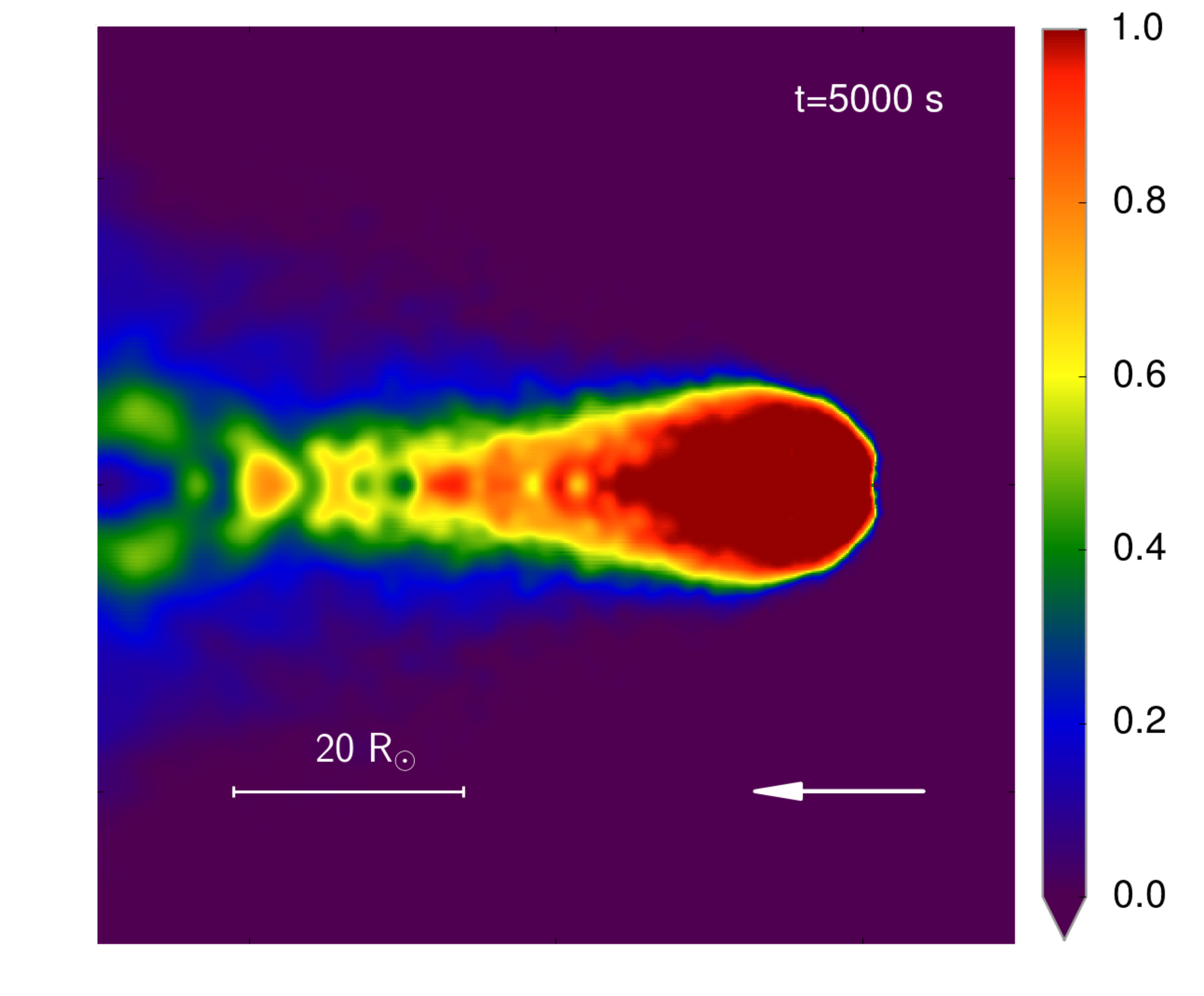}}
  \caption{ Mass fraction of companion material in the SN ejecta in our 
            simulation of the G/K-dwarf companion star model with a binary separation of $5.48\,R_{\odot}$ ($=7.1\;R_2$). The purple end of the color scale corresponds to pure SN 
            ejecta material, while the red end of the color scale represents pure companion material. 
             }
\label{Fig:mix}
  \end{center}
\end{figure}

\begin{figure*}
  \begin{center}
    {\includegraphics[width=2.0\columnwidth, angle=360]{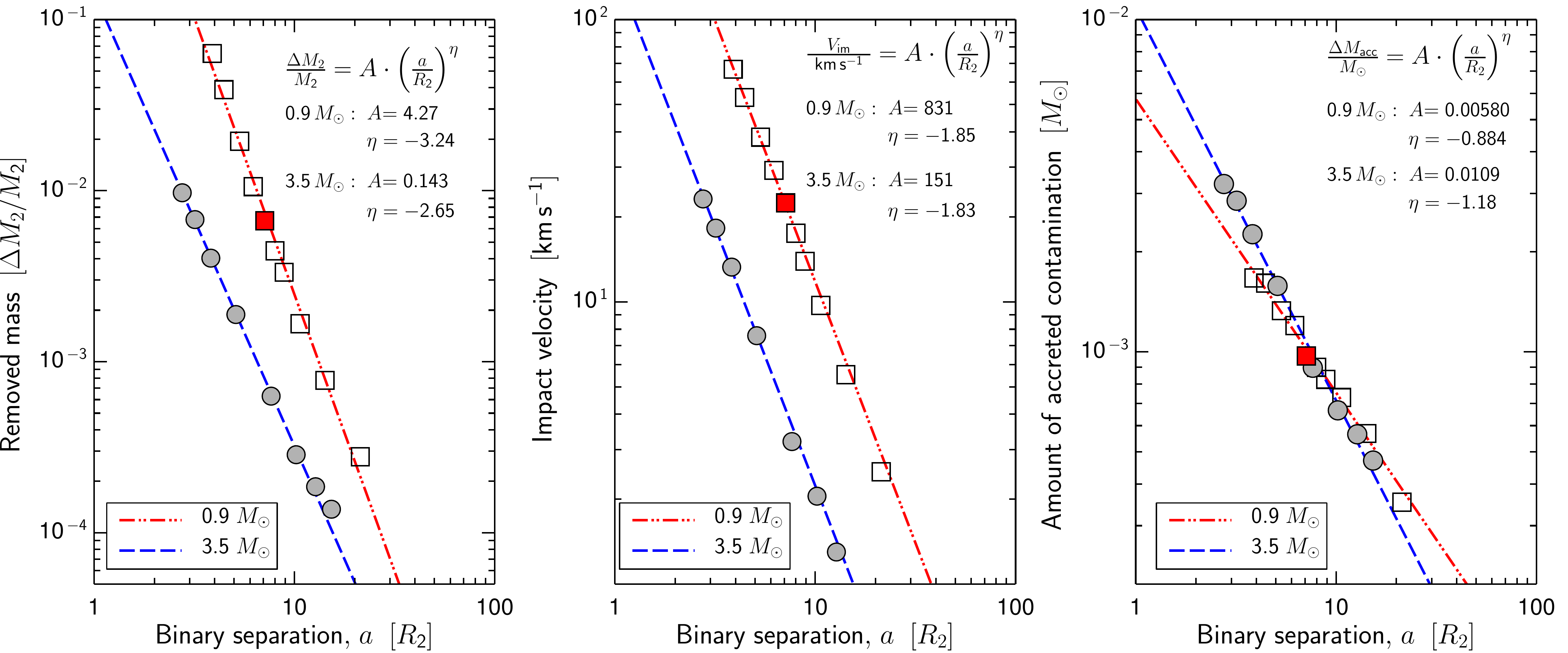}}
  \caption{Total removed companion mass (left panel), resulting impact velocity of the companion star (middle panel) and the amount of accreted contamination
           from the SN ejecta (right panel), as a function of initial binary separations for a G/K-dwarf (square symbols, $M_{2}=0.9\,M_{\odot}$ and $R_2\approx0.77\,R_{\odot}$) and 
           a late-type B-star (filled circle symbols, $M_{2}=3.5\,M_{\odot}$ and $R_2\approx2.18\,R_{\odot}$) companion model. The 
           results of our example model (Section~\ref{sec:results}) are shown with red filled 
           square symbols. Power-law fits
           are also  presented in each panel.}
\label{Fig:4}
  \end{center}
\end{figure*}

Despite the fact that the kinetic energy of the SN ejecta incident upon the companion star is much greater 
than the total binding energy of this star, the star survives the impact because energy is deposited in only
a small fraction of the stellar mass and excess energy is redirected into kinetic energy of the expelled material \citep{fa81,tf84}.   
However, the companion star receives linear momentum and hence a resulting impact velocity because of the collision with the SN ejecta, 
causing the star to move in the same direction as the SN ejecta with a speed of $\approx\! 22\,\rm{km\,s^{-1}}$. Furthermore, the surviving companion star is forced  
out of thermal equilibrium by internal heating from the passing shock wave. Hence it is expected that
it remains bloated on a thermal timescale (up to tens of Myr) after the SN explosion (cf. Section~\ref{sec:survivor}).  
To illustrate how the supernova ejecta are mixed with the removed companion material after the SN explosion,  the 
amount of material that originally belonged to the companion star is shown relative to the total 
amount of material for the simulation of a G/K-dwarf companion star model in Fig.~\ref{Fig:mix}. 
It is evident that most of the removed companion material is confined
to the downstream region behind the companion star, creating a hole in the SN ejecta with an opening 
angle of $\approx30^{\circ}$ with respect to the $x$-axis in our simulation.

Qualitatively, the impact proceeds in the same way in all our simulations. The numbers given above are derived 
in the special case of a simulation using $E_{\rm ej}=1.0\times 10^{51}\;{\rm erg}$. However, as we demonstrate systematically 
below, increasing the explosion energy enhances the effects of the SN ejecta impact.     

\subsection{Impact velocity of the companion star}\label{sec:evaporation}    
The impact velocity kick (up to about $100\;{\rm km\,s}^{-1}$, cf. Section~\ref{sec:discussion}), which the companion star receives
as a consequence of absorption of linear momentum from the SN ejecta, is in the orbital plane and opposite to
the direction to the exploding star. Although this impact velocity is small compared to the pre-SN orbital velocity of the
companion star, it increases the probability of disrupting the binary system and, in that case, also contributes
(although with a relatively small amount) to the escape velocity of the ejected star \citep{Taur98,Taur15}.
In addition, this impact velocity affects the eccentricity of the binary system if it survives the SN explosion.

The symmetry-breaking effect of orbital motion is ignored in our impact simulations 
because the orbital and spin velocities of the companion stars are much lower than the typical 
ejecta expansion velocity of a few $10^4\,\rm{km\,s^{-1}}$. We expect that some small 
morphological differences, such as the post-impact density distribution of all removed 
companion material and SN ejecta, would be seen if the orbital motion of the companion star is included, but 
it should not affect the basic results presented in this work (see also \citealt{Liu13b}).

\subsection{Contamination from the SN ejecta}\label{sec:contamination}
After the interaction with the SN ejecta, the envelope of the companion star may be enriched by
heavy elements from the SN ejecta while part of its H-rich material is stripped off by the SN impact. As a consequence, a surviving
companion star may show unusual chemical signatures compared to a normal MS star, in case a significant amount of SN-ejecta material is accumulated onto it.
These chemical peculiarities, if detected, can potentially differentiate between SN-induced HVSs from HVSs ejected from the Galactic center \citep{hil88}
or from dense stellar clusters by dynamical interactions \citep{leo91}. Possibly even the explosion mechanism (thermonuclear vs CCSN) could be determined
for these chemically enriched HVSs \citep{spb+15}.  

In our hydrodynamical simulations, we trace at every timestep all bound particles that originally belonged 
to the SN ejecta after the explosion. It takes some time (about $4\,000\,\rm{s}$ in our typical G/K-dwarf companion model) for this captured 
ejecta material to settle onto the surface of the companion star. At the end of the simulations,
the amount of ejecta mass accreted onto the surface of the surviving companion star
remains at a (roughly) constant value. In our example model (Fig.~\ref{Fig:3}), we find a total ejecta mass of 
$\sim\! 10^{-3}\,M_{\odot}$ is accumulated onto the surface of the companion star at the end of 
the simulation compared to the removal of $5.5\times 10^{-3}\;M_{\odot}$ of H-rich material. 
Unfortunately, we cannot provide detailed information of the chemical elements of the SN ejecta 
material that accumulate onto the surface of the star. This is because in our simulations only a simple 
ejecta model is constructed according to Eqs.~(\ref{eq:1}) and (\ref{eq:2}), without any information about the
chemical composition the ejecta material.

From our simulations, we find that most of the contamination is attributed 
to particles with low expansion velocity in the SN ejecta, i.e. the innermost region of 
the SN ejecta model \citep{Liu13c}. This is explained by the lower kinetic energy of those particles 
which makes it easier to accrete onto the surface of the companion star after the momentum 
transfer. Therefore, we argue that the composition of the ejecta material that pollutes
the companion star is sensitive to the nuclear burning at the center of the explosion.  
Based on CCSN modeling (e.g., \citealt{Nomo06}), we expect a potentially observable chemical enrichment of 
$\rm{Fe}$, $\rm{Si}$, and possibly $\rm{Ca}$ 
(e.g., \citealt{spb+15, Saba15}), if mixing processes (e.g., \citealt{Stan07, Stan08}) are not efficient on a long timescale.
Simulations with a more detailed model of the exploding star are needed to address 
the issue of chemical contamination in a future study.

In addition, we point out that the SPH method suppresses hydrodynamical instabilities seen in grid-based simulations, which may affect 
the estimated contamination from SN ejecta. However, earlier impact simulations with the same SPH 
approach \citep{Liu13a} and 3D grid-based simulations \citep{Pan12} found the same order of 
magnitude in the amount of nickel contamination on a He star donor from SN Ia ejecta. This indicates that the degree of instability captured in our SPH method is sufficient
for our study, or these instabilities may not play an important role in determining the amount of 
accreted SN ejecta in the simulation. This, however, should be tested in more detailed future studies.

\section{Discussion}\label{sec:discussion}

The kinetic energy, mass of the SN ejecta, the separation between the exploding star and its companion, and the structure of
the companion star at the time of the explosion are expected to be four major physical parameters of the progenitor system that 
influence the dynamics of the SN ejecta impact on the companion star. 
Different binary systems evolve to various evolutionary stages and have distinct binary parameters when the SN explodes. 
Consequently, some properties of the companion star and the binary system differ significantly from our example models. 
For instance, 
 the binary separation ($a$), 
 the mass of the exploding star ($M_{\rm He}$, which leads to different ejecta masses),
 and the structure (mass, $M_2$, and radius, $R_{2}$, and thus density profile) of the companion star at the moment  
 of the SN explosion, are all expected to be different in various binary systems.

\begin{figure}
  \begin{center}
    {\includegraphics[width=0.97\columnwidth, angle=360]{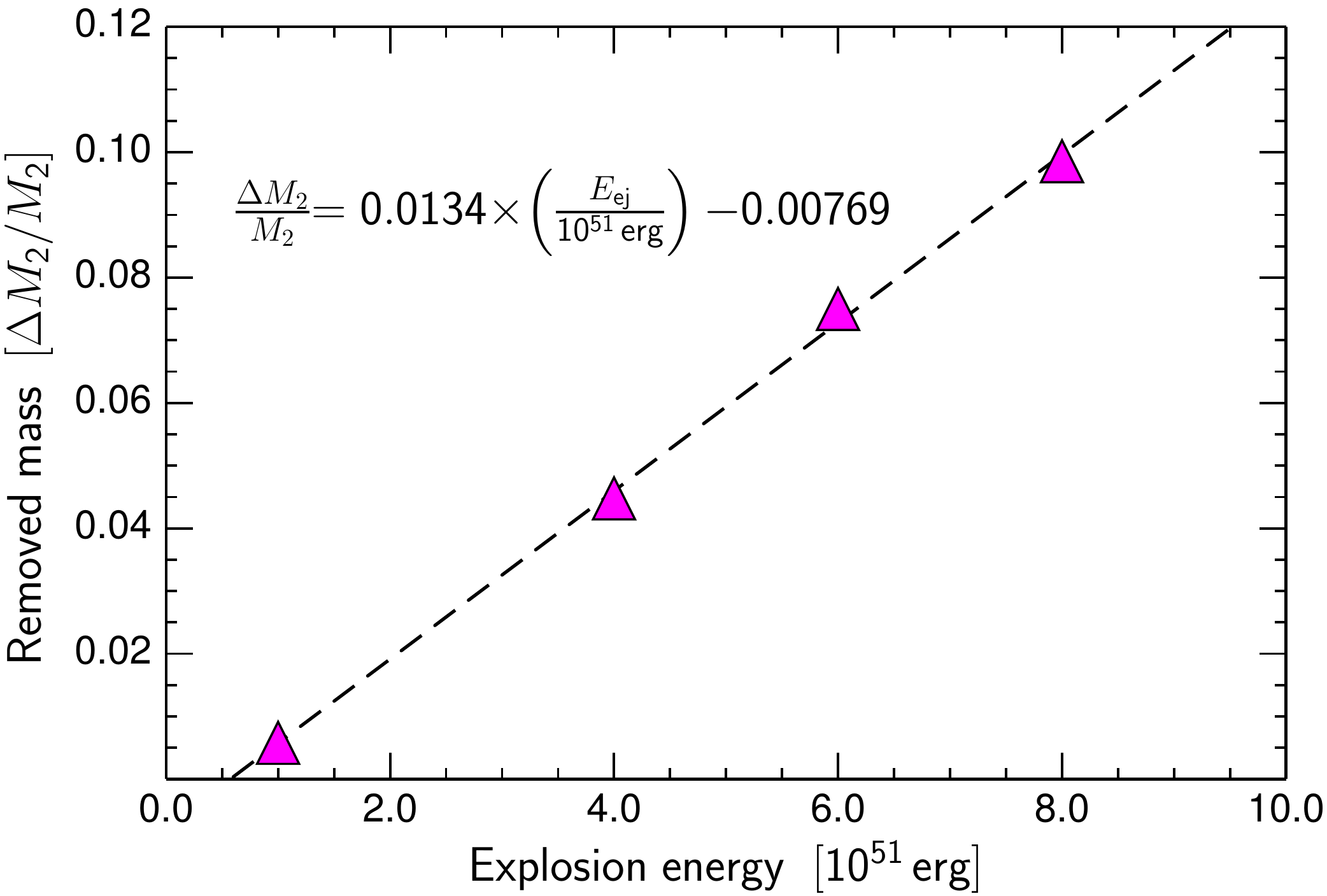}}
  \caption{Total removed companion mass ($\Delta M_2$, in a unit of $M_{2}$) as a function of the SN explosion energy for our $0.9\;M_{\odot}$ G/K-dwarf companion model. 
           All parameters except the SN explosion energy are kept constant, meaning that we only 
           change the SN explosion energy according to Eqs.~(\ref{eq:1}) and (\ref{eq:2}).
           Thus, for a fixed initial binary separation of $5.48\,R_{\odot}$ ($2\,a_{\rm min}$) and a total ejecta mass 
           of $1.4\,M_{\odot}$, different SN explosion energies of $(1.0,\,4.0,\,6.0,\,8.0)\times10^{51}\,\rm{erg}$ are
           investigated. The fitting function and fitting parameters are also shown.}
\label{Fig:5}
  \end{center}
\end{figure}

\begin{figure}
  \begin{center}
    {\includegraphics[width=0.97\columnwidth, angle=360]{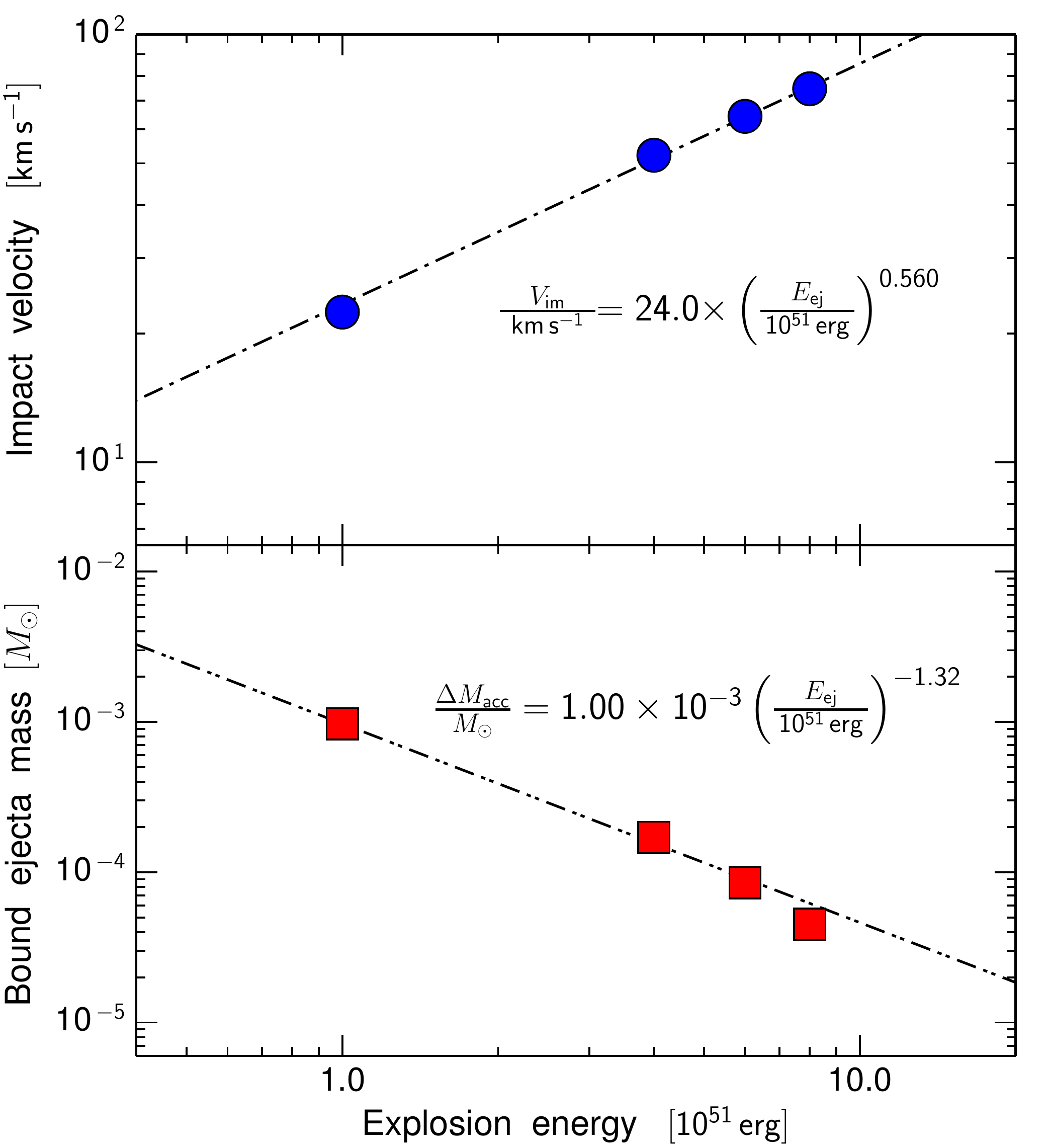}}
  \caption{As Fig.~\ref{Fig:5}, but for the resulting impact velocity ($v_{\rm im}$) and bound SN ejecta mass ($\Delta M_{\rm acc}$).}
\label{Fig:6}
  \end{center}
\end{figure}

\begin{figure}
  \begin{center}
    {\includegraphics[width=0.97\columnwidth, angle=360]{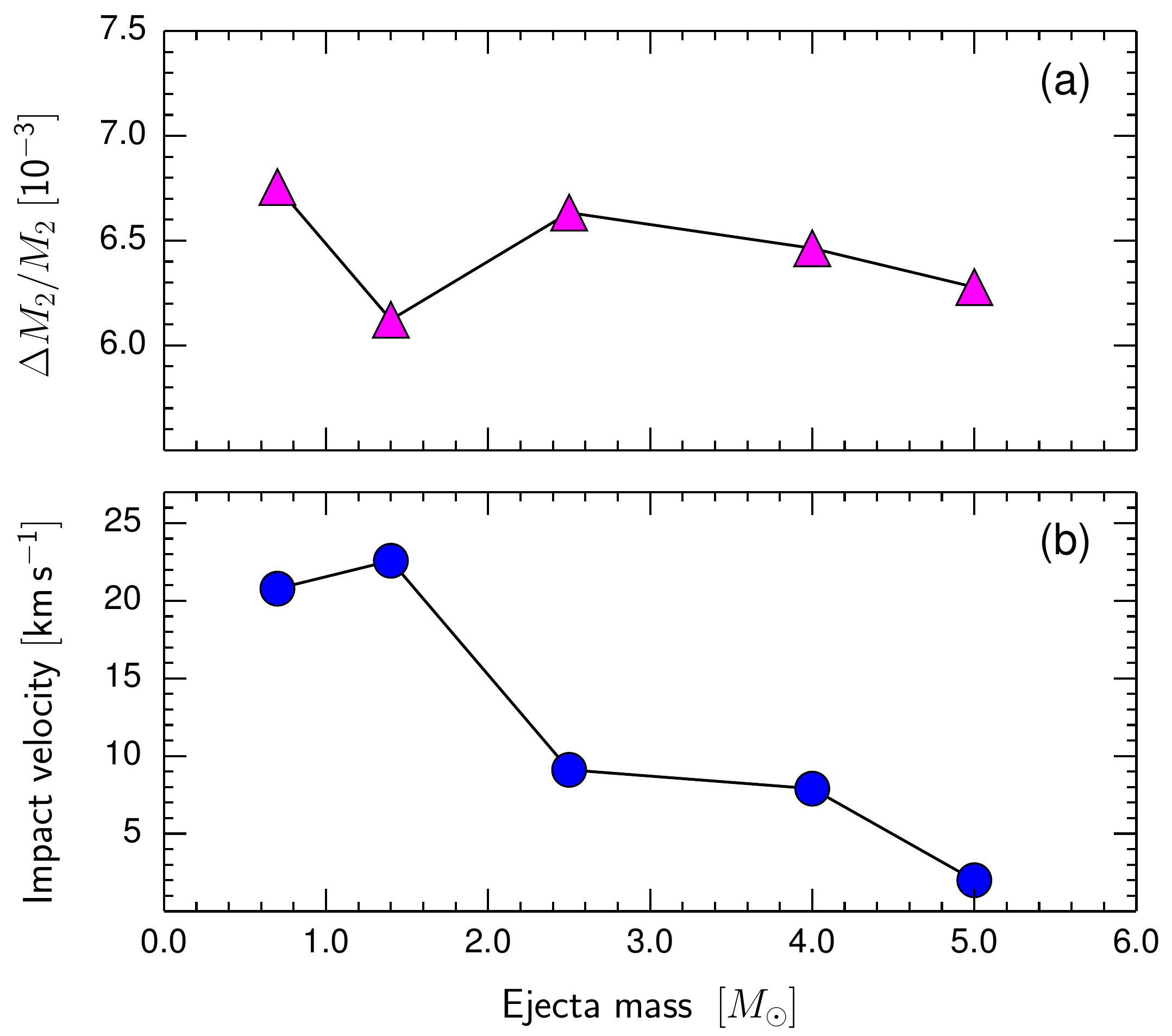}}
  \caption{As Figs.~\ref{Fig:5} and \ref{Fig:6}, but as a function of the SN ejecta mass. 
           All parameters (binary separation of $5.48\,R_{\odot}$ and a SN explosion energy 
           of $1.0\times10^{51}\,\rm{erg}$) except the SN ejecta mass ($M_{\rm ej}$) are kept constant. Different 
           values of $M_{\rm ej}=0.7,$ 1.4, 2.5, 4.0, and $5.0\,M_{\odot}$ are
           applied.  }
\label{Fig:7}
  \end{center}
\end{figure}

\begin{figure}
  \begin{center}
    {\includegraphics[width=0.97\columnwidth, angle=360]{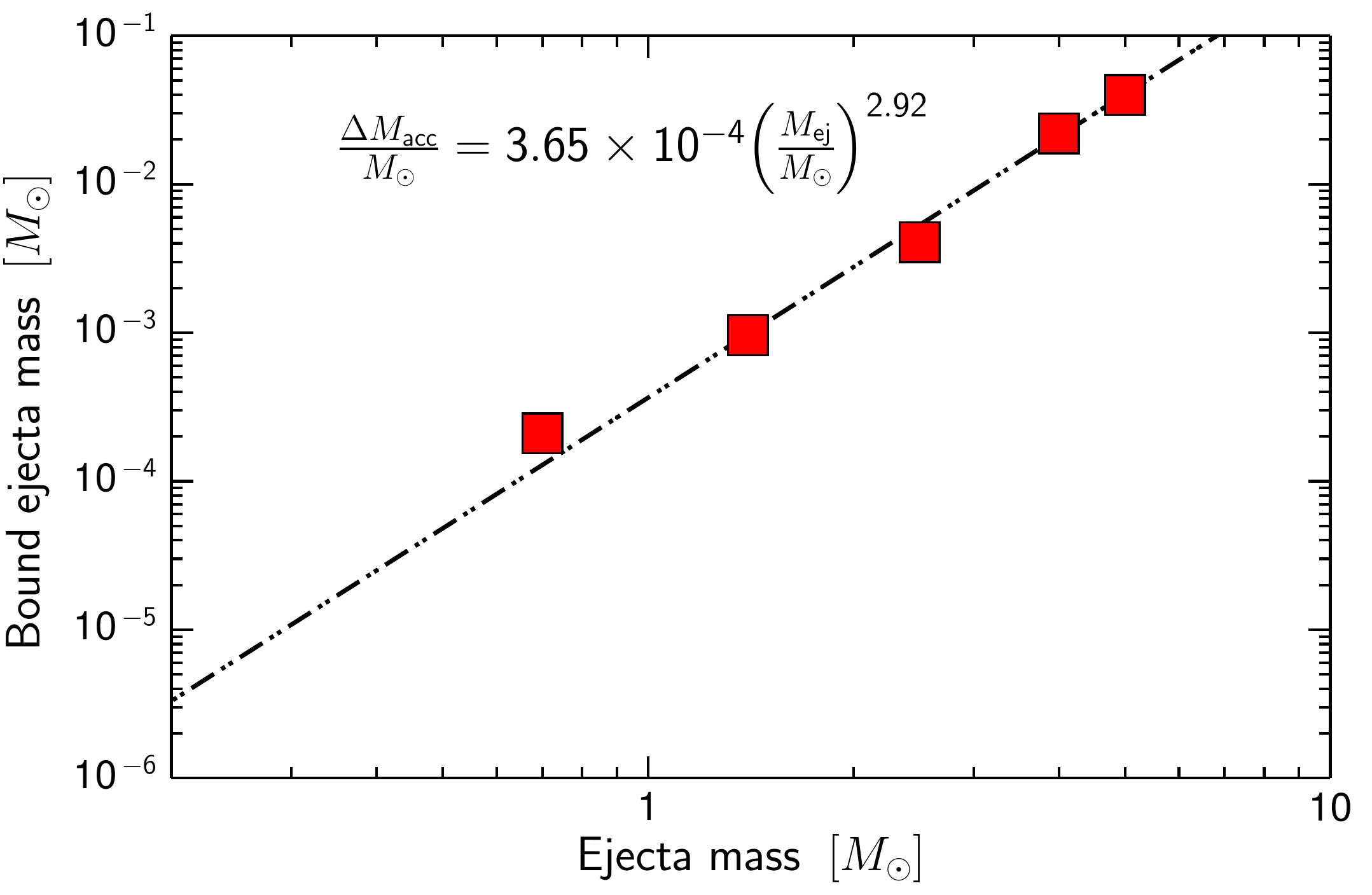}}
  \caption{As Fig.~\ref{Fig:7}, but for the amount of mass accreted from 
           the SN ejecta ($\Delta M_{\rm acc}$) as a function of the SN ejecta mass ($M_{\rm ej}$).}
\label{Fig:8}
  \end{center}
\end{figure}

In this section, we treat the binary separation ($a$) at the time of the explosion, the ejecta mass ($M_{\rm{ej}}$),  
or the SN explosion energy ($E_{\rm{ej}}$), as independent physical parameters, and we carry out a parameter survey 
to investigate the dependence of our numerical results on these parameters. Predictions from our population synthesis 
calculation are also discussed.

\subsection{Binary separation dependency}\label{sec:separation}
To test the influence of the ratio of binary separation to the companion star radius, $a/R_{2}$, we adjust the binary separation 
for a given companion star model. A fixed SN explosion model with an ejecta mass $M_{\rm{ej}}=1.4\,M_{\odot}$ 
and an explosion energy $E_{\rm{ej}}=1.0\times10^{51}\,\rm{erg}$ is used to model the SN event.

\begin{figure}
  \begin{center}
    {\includegraphics[width=0.98\columnwidth, angle=360]{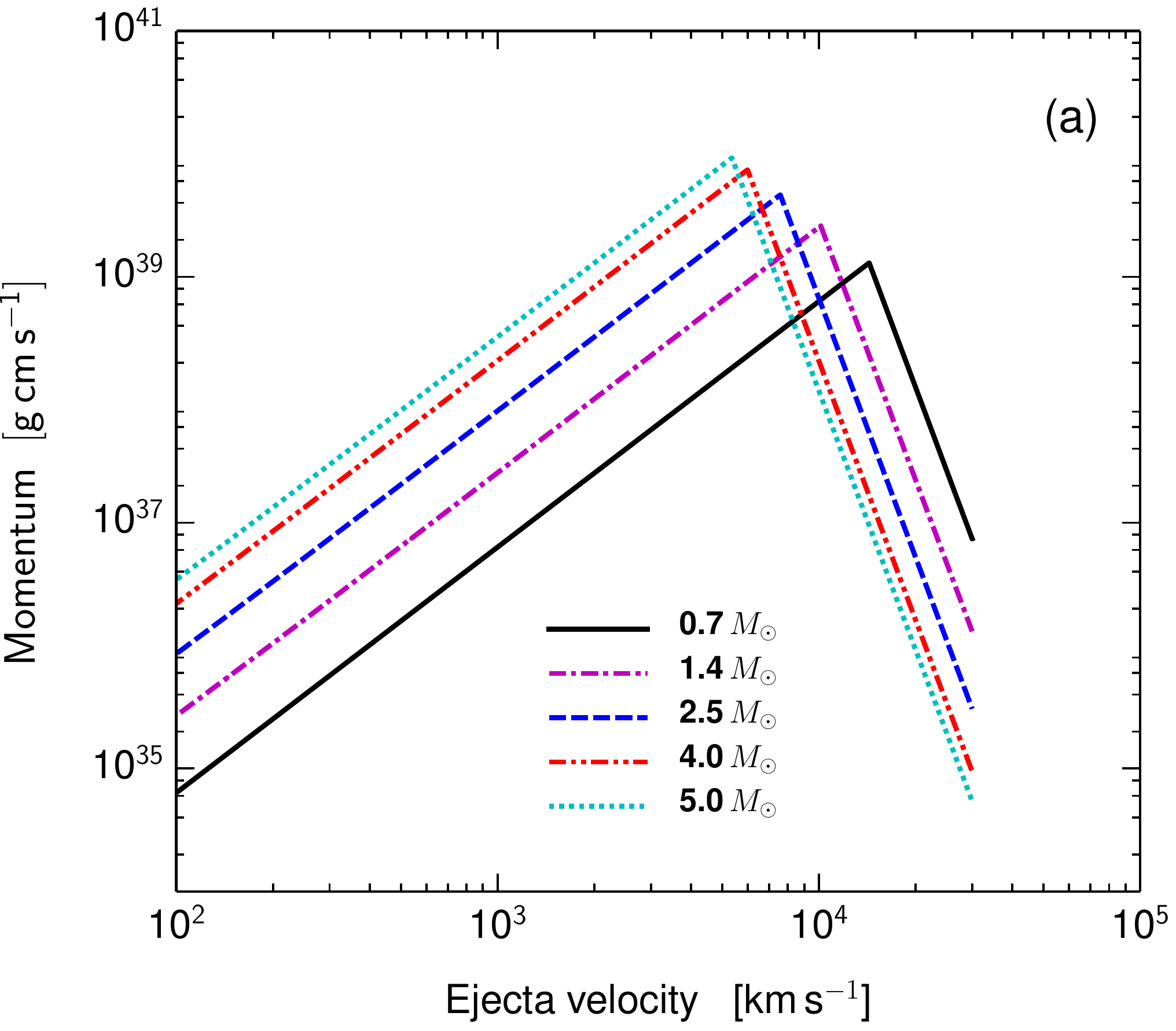}}
\hspace{0.2in}
    {\includegraphics[width=0.98\columnwidth, angle=360]{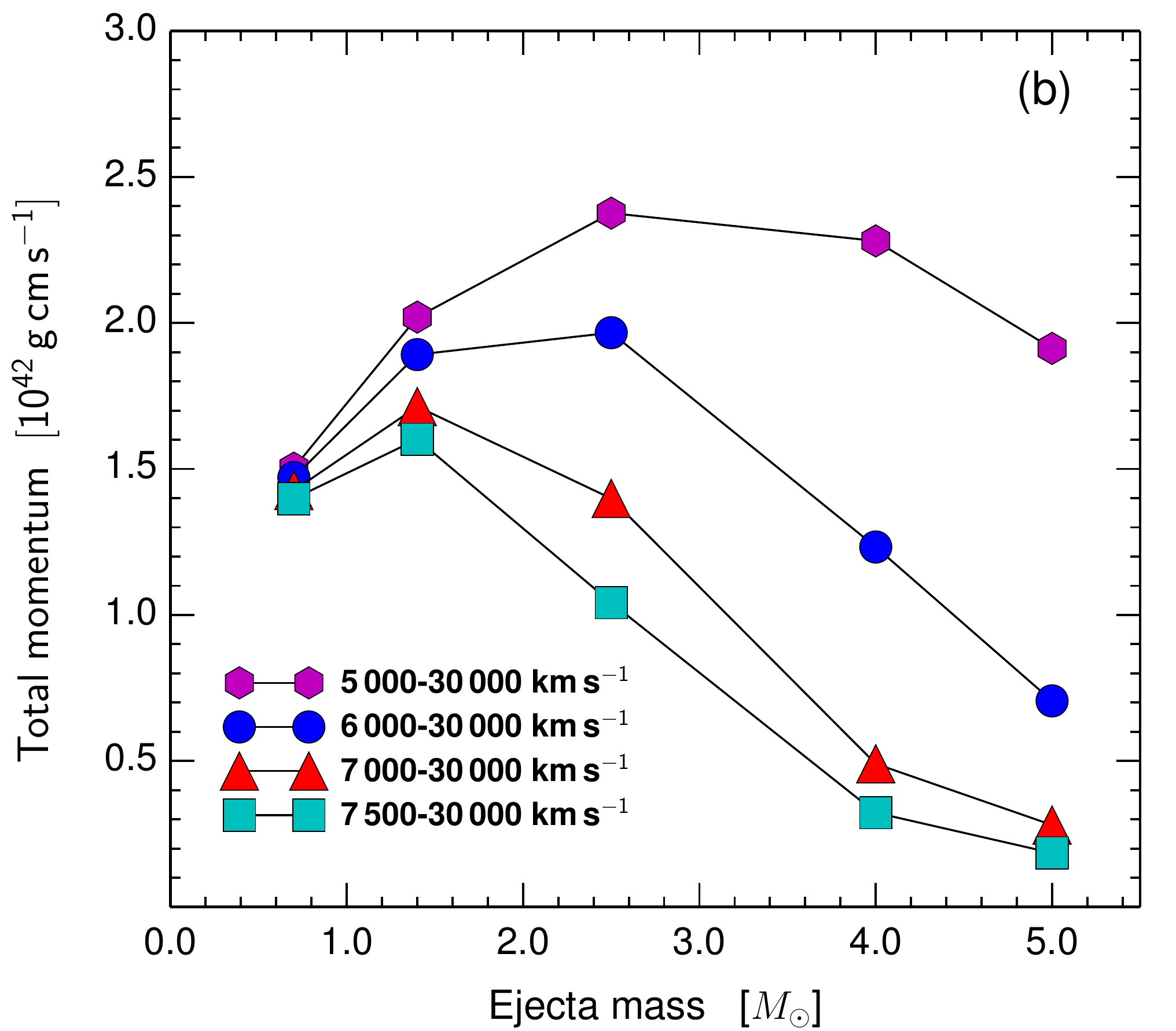}}
  \caption{Top: momentum within each velocity shell of initial SN ejecta as a function of ejecta velocity. 
           Bottom: total momentum of outer SN ejecta within regions with ejecta velocities $\gtrsim 5000$, $6000$, $7000$, and $7500\,\rm{km\,s^{-1}}$. Different ejecta 
           masses of $M_{\rm ej}=0.7$, 1.4, 2.5, 4.0 and $5.0\,M_{\odot}$ are
           applied.  }
\label{Fig:9}
  \end{center}
\end{figure}

Figure~\ref{Fig:4} shows the effects of varying the orbital separation parameter $a/R_{2}$, by a factor of about 6, 
on the total amount of removed companion mass ($\Delta M_2$), the resulting impact velocity ($v_{\rm im}$), 
and total accumulated ejecta mass ($\Delta M_{\rm acc}$), for the $0.9\,M_{\odot}$ and $3.5\,M_{\odot}$ 
companion star models. It shows that all three quantities 
decrease when increasing the orbital separation of the pre-SN binary. For instance, $\Delta M_2$ 
decreases by a factor of almost 10 as the parameter $a/R_{2}$ increases by 
a factor of 2. The relations can be approximately fitted with power-law functions (Fig.~\ref{Fig:4}),
which are consistent with previous impact simulations of thermonuclear (Ia) SNe (e.g., \citealt{Mari00,Pakm08,Liu12,Pan12}) and
CCSNe with a red giant companion star \citep{Hira14}.

A more careful inspection of Fig.~\ref{Fig:4} shows that the decrease in $\Delta M_2$ 
weakens systematically as the binary separation increases to be large, so systems with $a/R_{2}\ga8$ 
deviate slightly from the power law. This indicates that the detailed interactions 
are possibly more complex. For example, the amount of evaporated companion star material, and the accreted (bound) SN ejecta mass, may 
be poorly determined at long orbital separation because numerical difficulties were encountered when the removed mass 
(or accreted SN ejecta mass) is very small, i.e., when only a small number of SPH particles are involved. 
Furthermore, it cannot be ruled out that the impact simulation process also depends on the time evolution of the density of the 
SN ejecta hitting the star. 
However, we do not expect these uncertainties to strongly affect our basic results and conclusions, and 
the power-law fits presented here are fairly good approximations for the full range of $a/R_2$.  

\subsection{SN explosion energy dependency}\label{sec:energy}
The influence of the SN explosion energy on the ejecta interaction with the companion 
is studied on the basis of our G/K-dwarf companion star model with $a\approx5.48\,R_{\rm{\odot}} \approx 7.1\;R_2$. All parameters but 
the SN energy are kept constant.  
With a fixed ejecta mass of $M_{\rm ej}=1.4\;M_{\odot}$, we create four different SN ejecta models, using Eqs.~(\ref{eq:1}) and (\ref{eq:2}), 
with ejection energies (1.0--8.0)$\times10^{51}\,\rm{erg}$ \citep{Tadd15}.

Numerical results for the total removed companion mass, $\Delta M_2$, the resulting impact velocity, $v_{\rm im}$,
and total accreted mass from the SN ejecta (i.e., contamination), $\Delta M_{\rm acc}$,
are illustrated in Figs.~\ref{Fig:5} and \ref{Fig:6} as a function of the SN explosion energy, $E_{\rm ej}$.
The correlation between $\Delta M_2$ and $E_{\rm ej}$ is, to a good approximation, linear (Fig.~\ref{Fig:5}).
As expected, $\Delta M_2$ increases with $E_{\rm ej}$.
The behavior of both $v_{\rm im}$ and $\Delta M_{\rm acc}$ are better described by power-law functions (Fig.~\ref{Fig:6}).
Whereas $v_{\rm im}$ increases with $E_{\rm ej}$, $\Delta M_{\rm acc}$ is smaller at larger $E_{\rm ej}$.

\subsection{SN ejecta mass dependency}\label{sec:mass}
To investigate the influence of the SN ejecta mass, $M_{\rm ej}$, on the outcome of the SN impact, 
(i.e., $\Delta M_2$, $v_{\rm im}$ and $\Delta M_{\rm acc}$) 
we vary the SN ejecta mass by fixing all other parameters, 
i.e., $E_{\rm{ej}}=1.0\times10^{51}\,\rm{erg}$, $a=5.48\,R_{\odot}$, and apply a maximum 
ejecta velocity $\approx30\,000\,\rm{km\,s^{-1}}$. Under these assumptions, five SN ejecta models with ejecta 
masses 0.7, 1.4, 2.5, 4.0, and $5.0\,M_{\odot}$ are created using Eqs.~(\ref{eq:1}) and (\ref{eq:2}).

Figs.~\ref{Fig:7} and ~\ref{Fig:8} show the effect of the varying SN ejecta mass on the interaction between the SN 
ejecta and the companion star for our $0.9\;M_{\odot}$ G/K-dwarf companion model. With our method for creating 
different SN explosion profiles, we find that the SN ejecta does not significantly affect $\Delta M_2$ (Fig.~\ref{Fig:7}a).
However, it is found that the impact velocity of the companion star  
(Fig.~\ref{Fig:7}b) decreases nonmonotonically with increasing ejecta mass. 
This is a somewhat surprising and perhaps counterintuitive result given that 
for a fixed total kinetic energy, $E_{\rm{ej}}=1.0\times10^{51}\,\rm{erg}$, the incident momentum (i.e., the momentum 
of the SN ejecta that hits the star) is expected to increase by a factor of $2.7$ 
as the ejecta mass increases from 0.7 to $5.0\,M_{\odot}$ (using $E_{\rm ej}=p^2/(2M_{\rm ej})$). Therefore, the impact velocity of the star 
is expected to increase by a factor of about $2.7$ by the analytic analysis of \citet{Whee75}, 
which is, however, not seen in our simulations. There are several reasons for this
discrepancy between expected values from the analytic work and the results of our simulations. In 
Fig.~\ref{Fig:2}, it is shown that the impact velocity of the star almost reaches a maxium value about $600\,$--$700\,\rm{s}$ after
the SN explosion. This suggests that the impact velocity of the star is mostly determined by the 
momentum transfer during the collisions with high-velocity ejecta at early-time phase. 
Late-time interactions between the SN ejecta and the star do not significantly increase the impact velocity further because 
the drag on the companion star is significantly reduced at the late-time phase (\citealt{fa81}), and the momentum transfer 
due to ablation (which is dominant at the late-time phase) is quite inefficient. In addition, the shape of the companion star is changing 
while its mass is being removed by the SN impact, which may reduce its cross-sectional area and thus reduce the momentum 
transfer and drag efficiency \citep{fa81}. All these effects were not included in the analytic 
calculations of \citet{Whee75}. Therefore, the impact velocity of the companion star 
is very sensitive to the total momentum of the very outer SN ejecta layers with high expansion velocities. 
We refer to \citet{fa81} for further detailed descriptions of the impact velocity of the companion star after the SN impact.

In Figure~\ref{Fig:9}a, we present the momentum within each spherical velocity shell of the initial 
SN ejecta as a function of ejecta velocity for different ejecta mass models. In addition, the total momenta of the outer
SN ejecta regions with an expansion velocity in excess of $5\,000, 6\,000, 7\,000,$ and $7\,500\,\rm{km\,s^{-1}}$ are calculated for 
different ejecta mass models, respectively (Fig.~\ref{Fig:9}b). Our ejecta model leads to most of the 
momentum being distributed
in the inner part, i.e., low expansion velocity region, of the SN ejecta rather than in the outer part, 
i.e., high expansion velocity region, as  
we increase the SN ejecta mass (Fig.~\ref{Fig:9}a).
For example, as shown in Fig.~\ref{Fig:9}b, the total momentum within a region with expansion velocities larger 
than $7000\,\rm{km\,s^{-1}}$  increases first, then decreases with the increase of the SN ejecta mass. This high-velocity part 
of the SN ejecta hits the star 
within $\lesssim600\,\rm{s}$ after the explosion and is expected to be the main factor that determines
the final impact velocity of the companion star (Fig.~\ref{Fig:2}).  Therefore, it
is not surprising that the dependence of the impact velocity ($v_{\rm im}$) on the SN ejecta mass ($M_{\rm ej}$) in our simulations follows a pattern as shown
in Fig~\ref{Fig:7}b, keeping in mind that the impact velocity of the star strongly depends on the momentum distribution of 
the outer part of the SN ejecta. Adopting a SN ejecta model that is different from ours 
is therefore expected to yield a change in the resulting impact velocity of the star.

\begin{figure}
  \begin{center}
    {\includegraphics[width=0.95\columnwidth, angle=360]{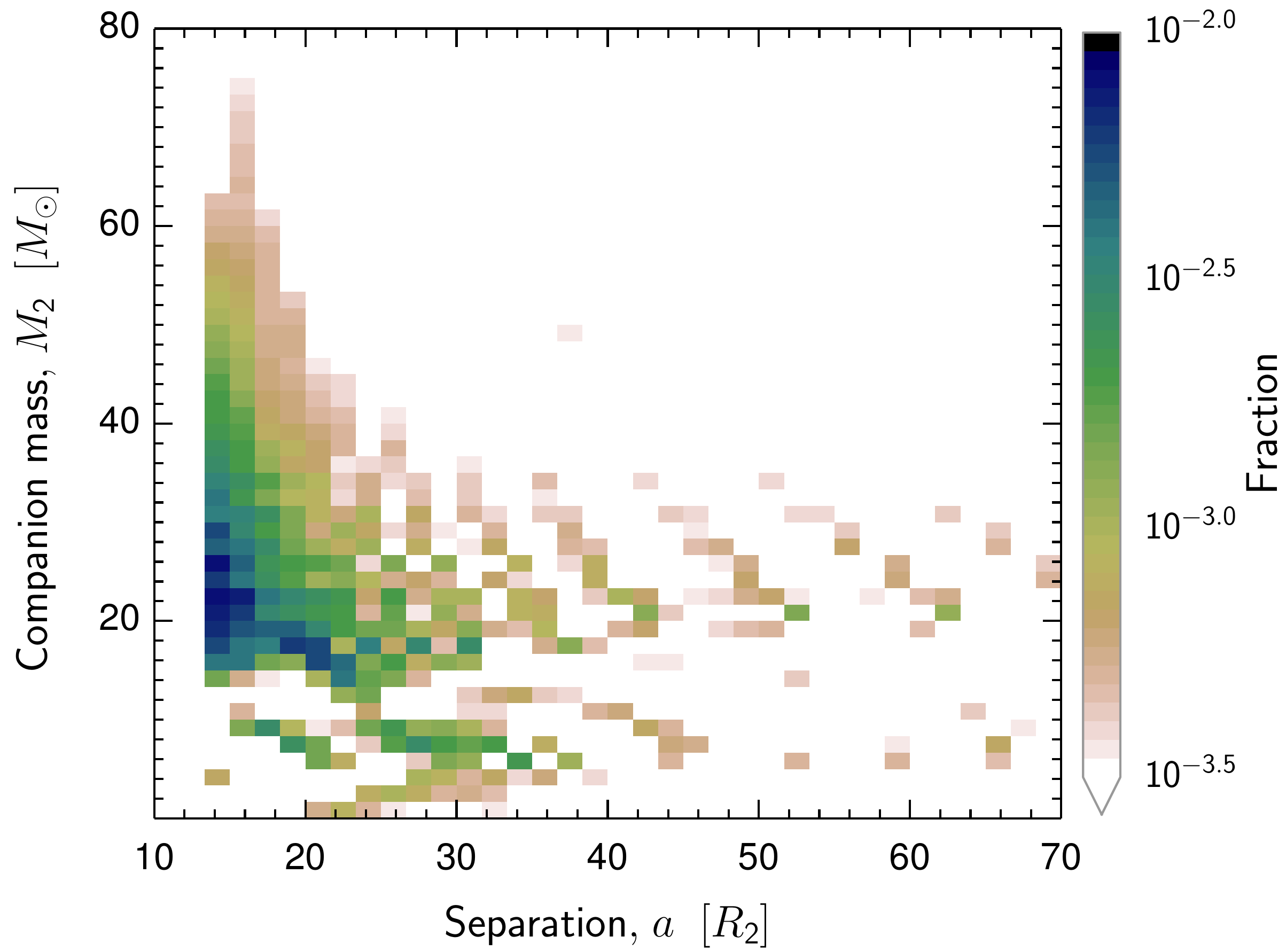}}
  \caption{Population synthesis distribution of the companion star mass ($M_{2}$) as a function of the binary separation ($a$) in 
          case the SN explodes as a Type Ib/c. Nothing is plotted in the regions with number fraction smaller than $10^{-3.5}$. 
 }
\label{Fig:10}
  \end{center}
\end{figure}

As expected, $\Delta M_{\rm acc}$ increases with larger $M_{\rm ej}$ and is approximately fitted by 
a power-law function (Fig.~\ref{Fig:8}). As already mentioned in Section~\ref{sec:contamination}, 
most of the ejected material captured by the companion star 
comes from the innermost low-velocity part of the SN ejecta (\citealt{Liu13a}). The enclosed mass of the
low-velocity ejecta increases as $M_{\rm ej}$ increases, which leads to more ejecta material being captured by the companion
star after the momentum transfer.

\begin{figure*}
  \begin{center}
    {\includegraphics[width=0.98\columnwidth, angle=360]{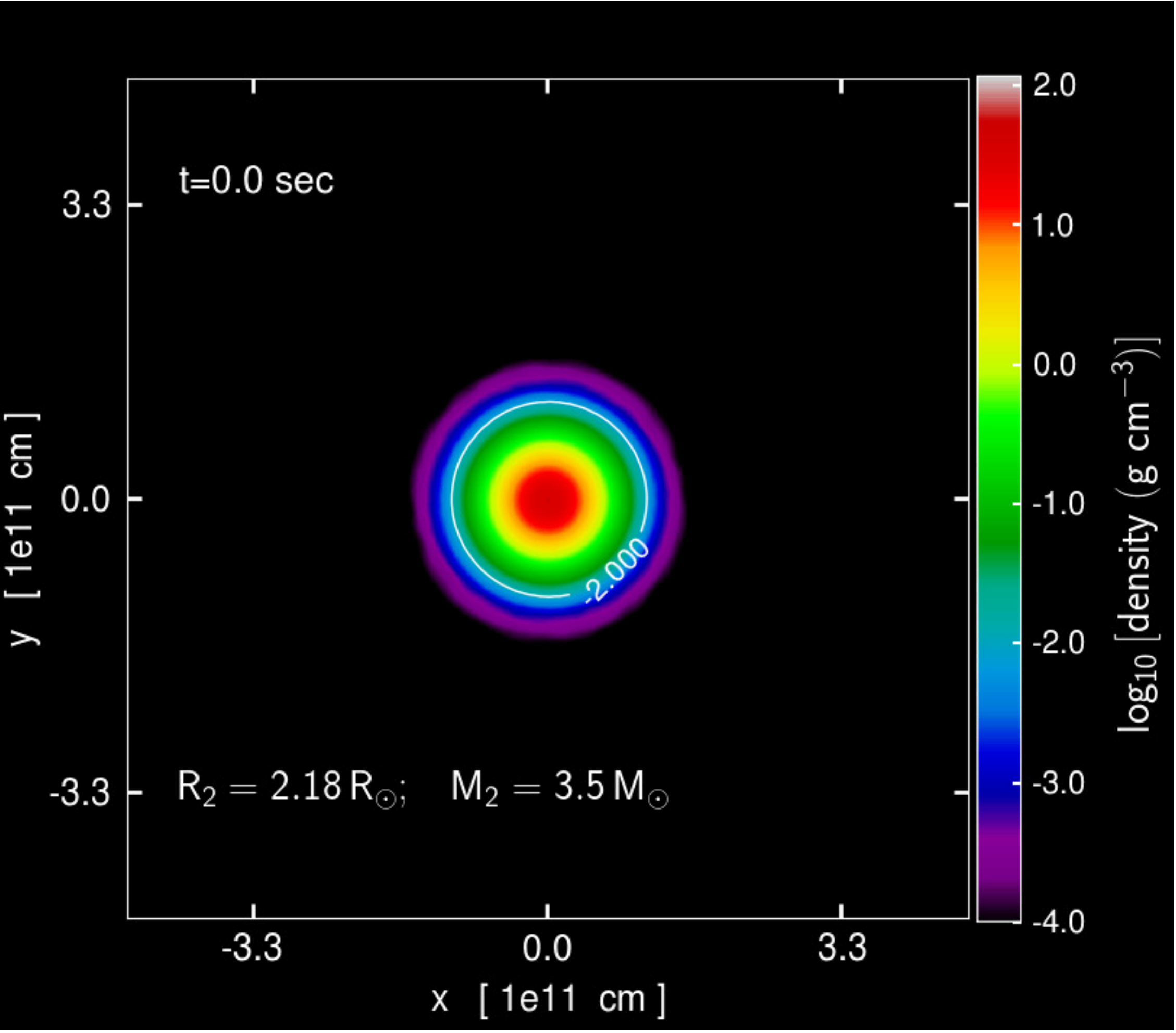}}
    {\includegraphics[width=0.98\columnwidth, angle=360]{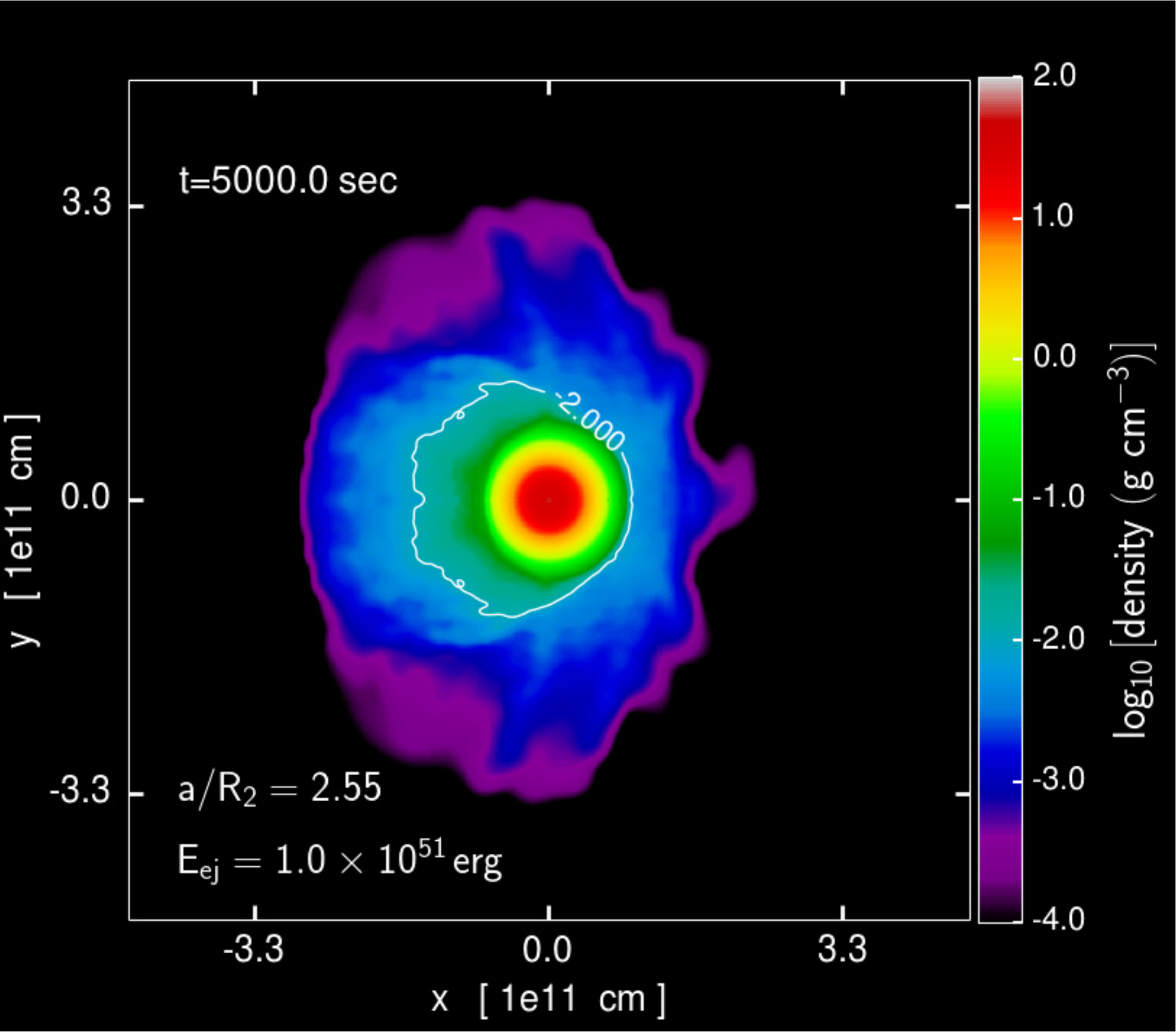}}
    {\includegraphics[width=0.98\columnwidth, angle=360]{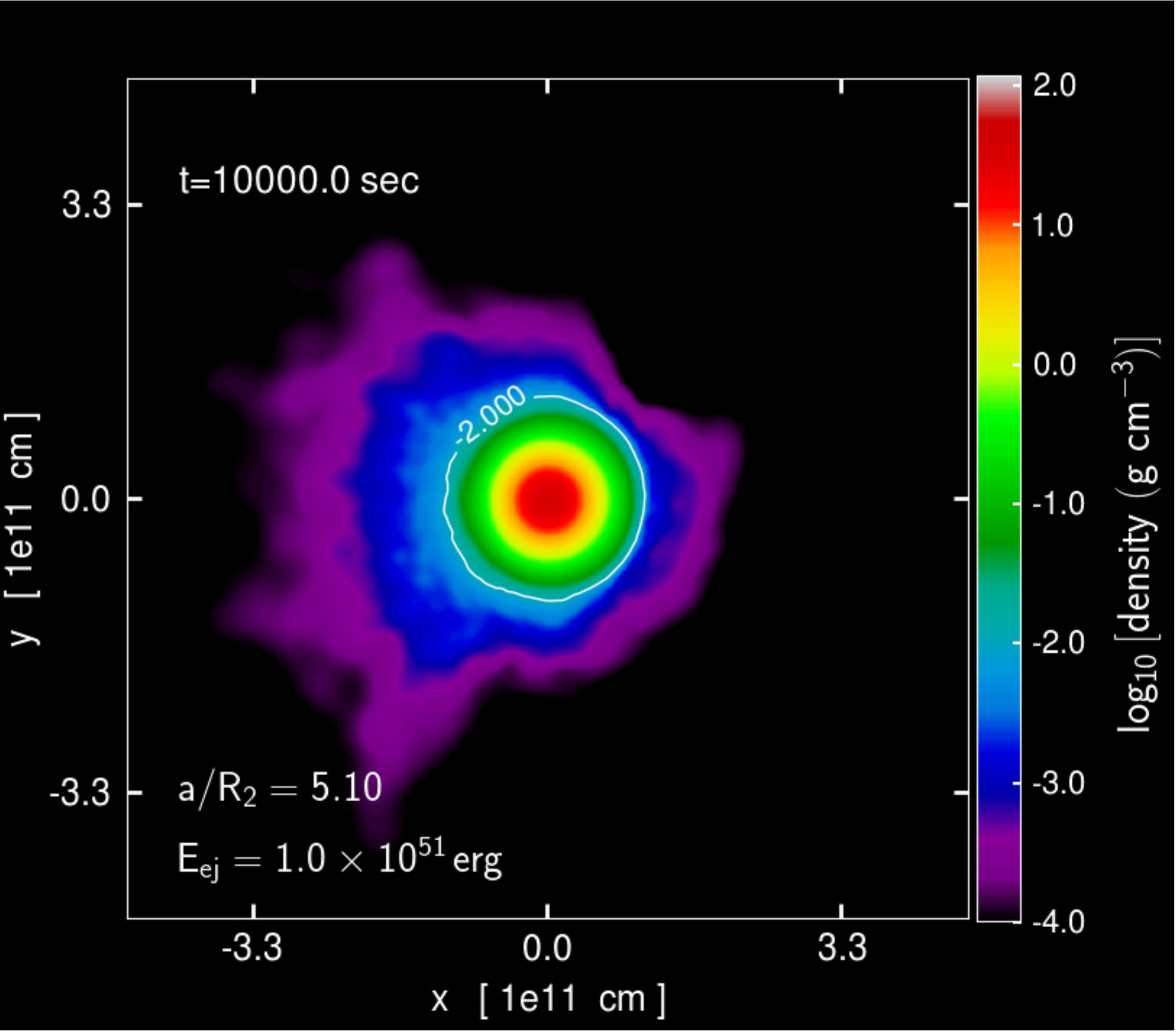}}
    {\includegraphics[width=0.98\columnwidth, angle=360]{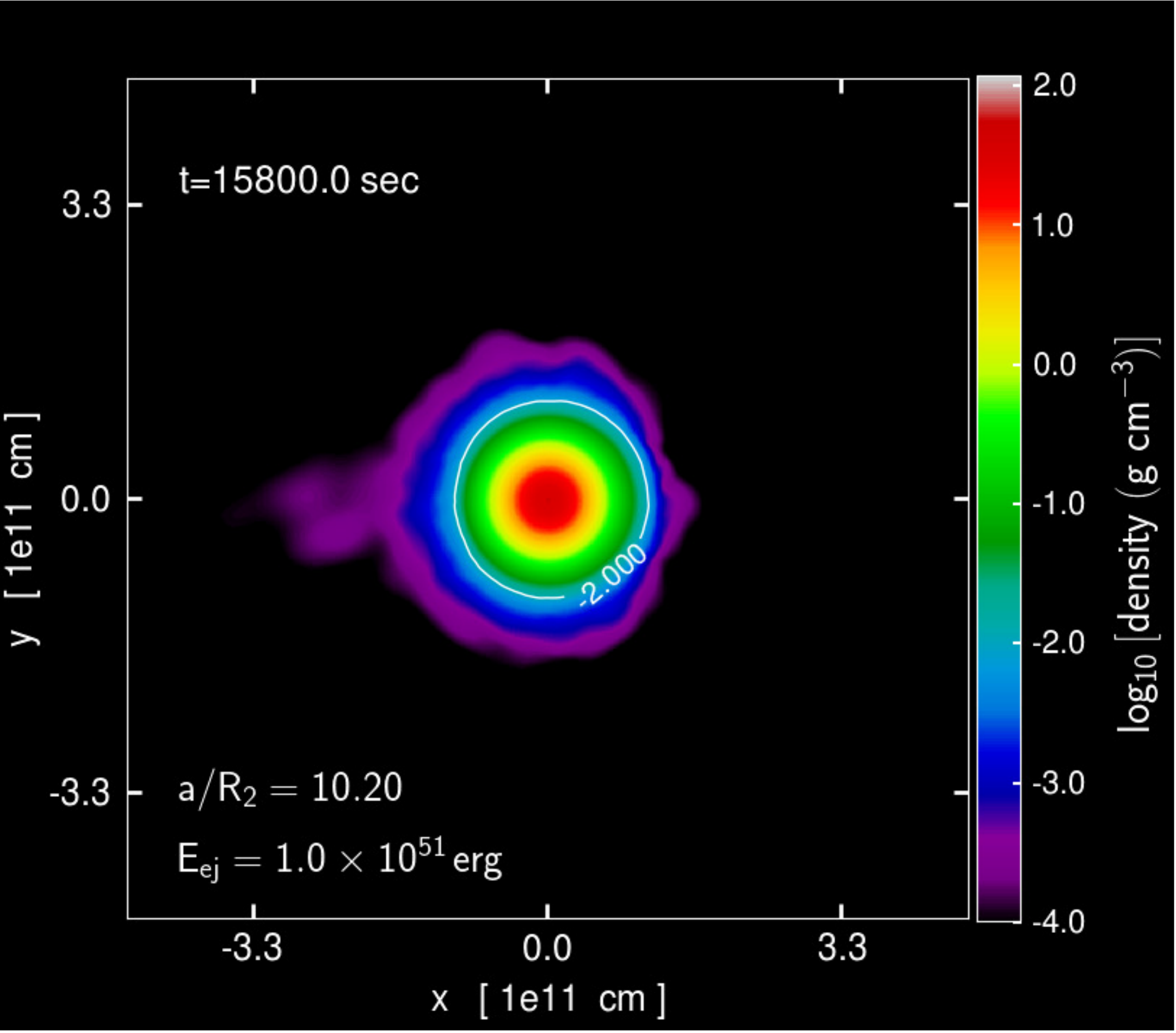}}
  \caption{ Density distribution simulations of the companion star using a $3.5\,M_{\odot}$ B-star MS model with different initial binary 
            separations. Only the bound material that originally belonged to the companion star and final accreted ejecta are shown. 
            The logarithm of density is color-coded. The top-left panel shows the initial configuration for all three cases.
            The motion of the incoming SN ejecta is from right to left. 
            The following three panels (for initial orbital separations of $a/R_2=2.55$, 5.10 and 10.20, respectively) 
            show snapshots of the mass density at the moment that the removed companion mass, the resulting impact velocity, 
            and the contamination from the SN ejecta, have converged to (almost) constant values. 
            In the widest system (lower-left panel), the impacted star is not significantly inflated compared to its initial state 
            before the SN impact.
            The white curves show constant density contours ($\rm{log_{10}\,\rho}=-2.0$). } 
\label{Fig:11}
  \end{center}
\end{figure*}

\subsection{Population synthesis predictions}\label{sec:bps}
To estimate the fraction of CCSNe in which light-curve brightening by the collision of SN ejecta with a stellar companion can be observed,
\citet{Mori15} perform populations synthesis calculations for SNe with the {\sc binary\_c/nucsyn} code \citep{Izza04, Izza06, Izza09}.
Under an assumption of a Galactic star formation rate of $0.68-1.45\,M_{\odot}\,\rm{yr^{-1}}$ and the average 
stellar mass of the Kroupa initial mass function ($0.83\,M_{\odot}$), the total CCSN rate of the 
Galaxy predicted from their population synthesis is $0.93$--$1.99\times10^{-2}\,\rm{yr^{-1}}$ (36\% 
SNe~Ib/c, 10\% SNe~IIb, 54\% SNe~II), 
consistent with the estimated Galactic CCSN rate of $2.30\pm0.48\times10^{-2}\,\rm{yr^{-1}}$ 
from recent surveys \citep{Li11}. In this section, we use the data from their population
synthesis calculations directly to discuss the effects of an explosion on the companion star in SNe~Ib/c. 
The basic parameters and assumptions in their calculations are summarized as follows.
The initial mass function of \citet{Krou01} is applied in combination with a flat distribution of mass ratios ($q=M_2/M_1$) and  
the assumption of circular binary orbits. The stars have a metallicity $Z=0.02$.
The orbital period distribution follows that of \citet{Sana12} for O-type primary stars in excess of $16\,M_{\odot}$,
\citet{Ragh10} for solar-like primary stars with a mass below $1.15\,M_{\odot}$ and a 
fitting function interpolated linearly in primary mass between these two limiting masses.
The common-envelope ejection efficiency is set to $\alpha_{\rm{CE}}=1$ and the envelope binding energy 
parameter, $\lambda$, is adapted from \citep{Dewi00,Dewi01,Taur01}. Systems with 
primary masses $3-80\,M_{\odot}$ and mass ratios between $0.1\,M_{\odot}/M_1$ and 1.0
are evolved with orbital periods between 0.1 and $10^{10}$ days.

Figure~\ref{Fig:10} presents the distributions  of the companion mass and the ratio of binary separation to companion radius
($a/R_{2}$) at the moment of the explosion for SNe~Ib/c. 
Most SNe~Ib/c 
have an orbital separation of $\gtrsim5.0\;R_2$, which is about a fraction of $\gtrsim95\%$ in our binary population synthesis calculations. 
Furthermore, with the distributions of $a/R_{2}$ in Fig.~\ref{Fig:10}, we can simply estimate $\Delta M_2$, $v_{\rm im}$ and $\Delta M_{\rm acc}$ 
by applying our power-law relationships stated in Fig.~\ref{Fig:4}.  
Typically, we find that less than $5\%$ of the companion mass is removed by the Type~Ib/c SN impact and the star 
receives an impact velocity of only a few $10\,\rm{km\,s^{-1}}$. In addition, typically a few $10^{-3}\,M_{\odot}$
of SN ejecta mass is found to accumulate onto its stellar surface after the SN impact. 

The above results are rough estimates based on the power-law relationships obtained from our simulations for a G/K-dwarf and 
a late-type~B companion star model (Fig.~\ref{Fig:4}). However, different fitting parameters are found for these two 
companion stars, which indicates that the detailed structures of the companion star could affect our numerical 
results. Based on previous simulations \citep{Liu12} the fitting parameters are expected to change slightly 
for different companion structures. Nevertheless, we do not expect 
significant discrepancies because, according to our population synthesis, a major fraction of the companion stars ($\gtrsim85\%$) are 
found to be in their MS phase when the Type~Ib/c SNe occur.

\subsection{Surviving companion stars}\label{sec:survivor}
After the SN explosion, the binary system may be disrupted. 
As mentioned in Sections~\ref{sec:density} and \ref{sec:contamination}, the companion stars surviving the SN explosion
probably exhibit some peculiar luminosity and chemical signatures and may thus be identified \citep[e.g.,][]{Pan12b} if 
these companion stars are strongly impacted and significantly heated by the SN ejecta.  

Figure~\ref{Fig:11} shows the mass density structures of surviving companion stars from our simulations 
of a late-type B-star companion model with different initial orbital separations. Again, we use the 
default SN ejecta model described in Section~\ref{sec:separation}. At small orbital separations, the 
companion star is strongly impacted and heated by the passing shock wave and hence it is bloated 
after the SN impact (top-right panel of Fig.~\ref{Fig:11}). As the orbital 
separation increases to $\approx\! 5\,R_{2}$, the SN impact and heating of the companion 
star are reduced. Only a small amount of companion mass is removed and the SN heating is too inefficient
to make the surviving companion star inflate by much (bottom row of Fig.~\ref{Fig:11}). 

Our population synthesis modeling suggests that most CCSNe ($\gtrsim95\%$) occur in systems 
with an orbital separation larger than about $5.0\,R_{2}$ (Section~\ref{sec:bps}). We caution that
there are large uncertainties in population synthesis studies, which may influence the results. 
In this case, these mainly relate to the input physics of common-envelope evolution and the subsequent Case~BB~RLO
from the naked helium star prior to its explosion \citep[][and references therein]{TvdH06}. 
In addition, Fig.~\ref{Fig:10} shows 
that at the moment of SN explosion most companion stars have a higher mass than our $3.5\,M_{\odot}$ late-type B-star companion. 
More massive stars have higher surface escape velocities and hence less companion mass 
is removed. Also, the fractional change in the internal stellar energy created by the shock wave is smaller because of
higher internal energy in more massive stars.
Consequently, if our population synthesis is correct, we do not expect that SN explosions 
significantly affect the detailed structures of companion stars in most CCSNe. This also implies that the effect on 
the post-impact evolution of the companion stars is not expected to be very dramatic. 
Therefore, some observable signatures, e.g., swelling and overluminosity, predicted for MS companions in SNe~Ia systems 
are, in general, not expected to be seen dramatically to the same extent in SNe~Ib/c and SNe~IIb systems. 
The difference is caused by the pre-SN orbital separations, which are, in general, smaller in SNe~Ia compared to SN~Ib/c systems. 
Simulations of SNe~Ia in the literature have been calculated with small binary separations, assuming that the companion stars fill their 
Roche-lobe radii at the time of the explosions \citep{Liu12, Pan12}.
Nevertheless, one should keep in mind that the ejected stars that become HVSs 
originate from the closest pre-SN orbits \citep{Taur15}, hence, SN-induced HVSs could 
suffer from SN ejecta impact effects. 

\citet{Hira15} studied SN ejecta-companion interactions for iPTF~13bvn and found that the companion star may expand extremely 
after the SN impact. They considered binary systems with a companion mass of $M_{2}\approx7.0-8.0\,M_{\rm{\odot}}$ and an orbital separation 
ratio of $a/R_{2}\approx20-30$. This result is different from the predictions for these wide binary systems in our simulations. However, it is difficult to 
discuss the reasons that cause these differences because the descriptions of detailed hydrodynamical impact simulations in \citet{Hira15} 
are limited.      
If the companion stars in binaries evolve to be red giant stars at the moment of CCSN explosion, 
the effect of the explosion is much greater because of the significantly reduced binding energy of 
the companion envelope (\citealt{Hira14}). However, as we have argued here, the probability that companion stars
are red giant stars at the moment of a SN explosion is relatively low ($<$~15\%). 

Finally, for fixed values of $E_{\rm ej}$ and $M_{\rm ej}$ the calibrated fitting formulae for the removed mass or impact velocity 
in Appendix~A of \citet{Taur15}, for a $1.0\;M_{\odot}$ MS star, 
are in good agreement with our hydrodynamical simulations for a $0.9\;M_{\odot}$ MS star, yielding similar
values for $\Delta M_2$ and $v_{\rm im}$.

\section{Summary and conclusions}\label{sec:summary}

We have investigated the impact of SN ejecta on the companion stars in CCSNe of Type~Ib/c using the SPH code {\sc Stellar GADGET}. 
We focused on 0.9 and $3.5\;M_{\odot}$ MS companion stars constructed with the {\sc BEC} stellar evolution code. 
Our main results can be summarized as follows: 

\begin{itemize}
\item[1.]  The dependence of total removed mass ($\Delta M_2$) and resulting impact velocity ($v_{\rm im}$) on the pre-SN binary separation ($a$)
           can be fitted with power-law functions, which is consistent with previous work on SNe~Ia. 
           The amount of accreted SN ejecta mass ($\Delta M_{\rm acc}$) is found to be dependent on $a$ with a power-law relationship.
           All three quantities are shown to decrease significantly with increasing $a$, as expected.
\item[2.]  The relation between $\Delta M_2$ and the SN explosion energy ($E_{\rm ej}$) is linear. 
           The quantities $v_{\rm im}$ and $\Delta M_{\rm acc}$ can better be fitted with 
           power-law functions and their values increase and decrease, respectively, 
           with larger values of $E_{\rm ej}$. 
\item[3.]  With a fixed value of $E_{\rm ej}$, the amount of SN ejecta mass ($M_{\rm ej}$) is found to not affect $\Delta M_2$ by much.  
           However, more ejecta mass can be accumulated onto the companion star in simulations
           with larger value of $M_{\rm ej}$ for a fixed value of $E_{\rm ej}$.
           The impact velocity, $v_{\rm im}$, is sensitive to the momentum profile of the outer SN ejecta. With our SN ejecta model, 
           we find that the impact velocity, $v_{\rm im}$, decreases at larger $M_{\rm ej}$.
\item[4.]  Our population synthesis calculations predict that in most CCSNe less than $5\%$ of 
           the MS companion mass can be removed by the SN impact 
           (i.e., $\Delta M_2/M_2<0.05$). In addition, the companion star typically receives an impact velocity, $v_{\rm im}$,
           of a~few $10\,\rm{km\,s^{-1}}$, and the amount of SN ejecta captured by the companion star after the explosion, $\Delta M_{\rm acc}$, 
           is most often less than $10^{-3}\,M_{\odot}$. 
\item[5.]  Because a typical CCSN binary companion is relatively massive and can be located at a large pre-SN distance,
           we do not expect, in general, that the effects of the SN explosion on the post-impact stellar evolution will be very dramatic.
\item[6.]  In the closest pre-SN systems, the MS companion stars are affected more strongly by the SN ejecta impact, leading to
           $\Delta M_2/M_2\simeq 0.10$, $v_{\rm im}\simeq 100\;{\rm km\,s}^{-1}$ and $\Delta M_{\rm acc}\simeq 4\times 10^{-3}\;M_{\odot}$, depending
           on the mass of the companion star. In addition, these stars are significantly bloated as a consequence of internal heating by
           the passing shock wave.
\item[7.]  It is possible that the SN-induced high velocity stars (HVSs), or more ordinary, less fast, runaway stars, 
           may be sufficiently contaminated to be identified
           by their chemical peculiarity as former companion stars to an exploding star if mixing processes are not 
           efficient on a long timescale.  

\end{itemize}

\section*{Acknowledgements}
We thank the anonymous referee for his/her valuable comments and suggestions that helped to improve 
the paper. We would like to thank Carlo Abate for helpful discussions. This work is supported by the Alexander von Humboldt Foundation.
T.M.T. acknowledges the receipt of DFG Grant: TA 964/1-1. 
The work of F.K.R. is supported by the ARCHES prize of the German Federal Ministry of Education and Research (BMBF). 
T.J.M. is supported by Japan Society for the Promotion of Science
Postdoctoral Fellowships for Research Abroad (26\textperiodcentered 51).
R.J.S. is the recipient of a Sofja Kovalevskaja Award from the Alexander von Humboldt Foundation.      
R.G.I. thanks the AvH and Science and Technologies Research Council UK for his Rutherford Fellowship.
Z.W.L. acknowledges the computing time granted by the Yunnan Observatories
and provided on the facilities at the Yunnan Observatories
Supercomputing Platform. The simulations were partially carried out using Supercomputing Platform of
the Yunnan Observatories, Kunming, People’s Republic of China.

\bibliographystyle{aa}

\bibliography{ref}

\end{document}